\documentclass{aa}  

\usepackage{graphicx}
\usepackage{txfonts}
\usepackage{multirow}
\usepackage{colortbl}

\usepackage{hyperref}  
\hypersetup{colorlinks=true,linkcolor=[rgb]{1.,0.2,0.2},citecolor=[rgb]{0.1,0.4,1.},filecolor=[rgb]{0.7,0.2,0.2},urlcolor=[rgb]{0.7,0.2,0.2}}

\usepackage{color}
\definecolor{blue}{rgb}{0., 0., 1}

\newcommand {\EW}{EW$_0$}

\newcommand{\hi}{\textrm{H}\textsc{i}}

\newcommand{\oiii}{[\textrm{O}\textsc{iii}]}
\newcommand{\oii}{[\textrm{O}\textsc{ii}]}

\newcommand{\oiidoublam}{[\textrm{O}\textsc{ii}]\ensuremath{\lambda3727,3729}}
\newcommand{\oiilam}{[\textrm{O}\textsc{ii}]\ensuremath{\lambda3727}}

\newcommand{\oiiiv}{[\textrm{O}\textsc{iii}]\ensuremath{\lambda5007}}
\newcommand{\oiiiiv}{[\textrm{O}\textsc{iii}]\ensuremath{\lambda4959}}

\newcommand{\oiiidoublam}{[\textrm{O}\textsc{iii}]\ensuremath{\lambda\lambda4959,5007}}
 
\newcommand{\ha}{\ifmmode {\rm H}\alpha \else H$\alpha$\fi}
\newcommand{\halam}{\ifmmode {\rm H}\alpha \lambda6563 \else H$\alpha$ $\lambda$6563 \fi}
\newcommand{\hb}{\ifmmode {\rm H}\beta \else H$\beta$\fi}
\newcommand{\hg}{\ifmmode {\rm H}\gamma \else H$\gamma$\fi}
\newcommand{\hblam}{\ifmmode {\rm H}\beta \lambda4861 \else H$\beta$ $\lambda$4861 \fi}
\newcommand{\lya}{\ifmmode {\rm Ly}\alpha \else Ly$\alpha$\fi}
\newcommand{\pg}{\ifmmode {\rm P}\gamma \else Pa$\gamma$\fi}
\newcommand{\lyb}{\ifmmode {\rm Ly}\beta \else Ly$\beta$\fi}
\newcommand{\lyg}{\ifmmode {\rm Ly}\gamma \else Ly$\gamma$\fi}

\newcommand{\civ}{\textrm{C}\textsc{iv}\ensuremath{\lambda1548,1550}}

\newcommand{\heii}{\textrm{He}\textsc{ii}\ensuremath{\lambda1640}}
\newcommand{\heiiopt}{\textrm{He}\textsc{ii}\ensuremath{\lambda4686}}

\newcommand{\flyc}{\ifmmode  \mathrm{f}_\mathrm{esc}\mathrm{(LyC)} \else $\mathrm{f}_\mathrm{esc}\mathrm{(LyC)}$\fi}

\def\sfr{M$_{\odot}$~yr$^{-1}$}
\def\kms{km s$^{-1}$}

\def\ergs{\ifmmode \mathrm{erg\hspace{1mm}s}^{-1} \else erg s$^{-1}$\fi}
\def\ergscm{erg s$^{-1}$ cm$^{-2}$}
\def\micron{\ifmmode \mu\mathrm{m} \else $\mu$m\fi}
\def\msun{\ifmmode \mathrm{M}_{\odot} \else M$_{\odot}$\fi}
\def\msunyr{\ifmmode \mathrm{M}_{\odot} \hspace{1mm}{\rm yr}^{-1} \else $\mathrm{M}_{\odot}$ yr$^{-1}$\fi}
\def\zsun{\ifmmode Z_{\odot} \else Z$_{\odot}$\fi}
\def\lsun{\ifmmode L_{\odot} \else L$_{\odot}$\fi}
\def\mstar{\ifmmode \mathrm{M}_{\star} \else M$_{\star}$\fi}
\newcommand{\JWST}{\textit{JWST}}
\newcommand{\HST}{\textit{HST}}
\newcommand{\lap}{\textit{LAP1}}

\usepackage{orcidlink}

\newcommand{\orcid}[1]{\href{https://orcid.org/#1}{\textcolor[HTML]{A6CE39}{\aiOrcid}}}

\begin{document}

\titlerunning{\JWST\ probes metal poor stellar complex at $z\simeq6.64$}
\title{An extremely metal poor star complex in the reionization era: Approaching Population III stars with JWST
\thanks{Based on observations collected with the James Webb Space Telescope (\JWST) and Hubble Space Telescope (\HST). 
These observations are associated with \JWST\ GO program n.1908 (PI E. Vanzella) and GTO n.1208 (CANUCS, PI C. Willot).}}

\authorrunning{Eros Vanzella et al.}
\author{
E.~Vanzella\inst{\ref{inafbo}}\fnmsep\thanks{E-mail: \href{mailto:eros.vanzella@inaf.it}{eros.vanzella@inaf.it}}$^{\orcidlink{0000-0002-5057-135X}}$,
F.~Loiacono\inst{\ref{inafbo}}$^{\orcidlink{0000-0002-8858-6784}}$,
P.~Bergamini \inst{\ref{unimi},\ref{inafbo}}$^{\orcidlink{0000-0003-1383-9414}}$,
U.~Me\v{s}tri\'{c} \inst{\ref{unimi},\ref{inafbo}}$^{\orcidlink{0000-0002-0441-8629}}$,
M.~Castellano\inst{\ref{inafroma}}$^{\orcidlink{0000-0001-9875-8263}}$,
P.~Rosati \inst{\ref{unife},\ref{inafbo}}$^{\orcidlink{0000-0002-6813-0632}}$,
M.~Meneghetti \inst{\ref{inafbo}}$^{\orcidlink{0000-0003-1225-7084}}$,
C.~Grillo \inst{\ref{unimi},\ref{inafiasf}}$^{\orcidlink{0000-0002-5926-7143}}$,
F.~Calura\inst{\ref{inafbo}}$^{\orcidlink{0000-0002-6175-0871}}$,
M.~Mignoli\inst{\ref{inafbo}}$^{\orcidlink{0000-0002-9087-2835}}$,
M.~Brada{\v c}\inst{\ref{uniLjubljana},\ref{unicalifornia}}$^{\orcidlink{0000-0001-5984-0395}}$,
A.~Adamo\inst{\ref{univstock}}$^{\orcidlink{0000-0002-0786-7307}}$,
G.~Rihtar{\v s}i{\v c}\inst{\ref{uniLjubljana}}$^{\orcidlink{0009-0009-4388-898X}}$,
M.~Dickinson\inst{\ref{tucson}}$^{\orcidlink{0000-0001-5414-5131}}$,
M.~Gronke\inst{\ref{maxplanck}}$^{\orcidlink{0000-0003-2491-060X}}$,
A.~Zanella \inst{\ref{inafpd}}$^{\orcidlink{0000-0001-8600-7008}}$,
F.~Annibali\inst{\ref{inafbo}}$^{\orcidlink{0000-0003-3758-4516}}$,
C.~Willott\inst{\ref{nrccanada}}$^{\orcidlink{0000-0002-4201-7367}}$,
M.~Messa\inst{\ref{univstock},\ref{unigeneva}}$^{\orcidlink{0000-0003-1427-2456}}$,
E.~Sani\inst{\ref{esochile}}$^{\orcidlink{0000-0002-3140-4070}}$,
A.~Acebron\inst{\ref{unimi}}$^{\orcidlink{0000-0003-3108-9039}}$,
A.~Bolamperti\inst{\ref{inafpd},\ref{inafbo},\ref{eso_germany}}$^{\orcidlink{0000-0001-5976-9728}}$,
A.~Comastri\inst{\ref{inafbo}}$^{\orcidlink{0000-0003-3451-9970}}$,
R.~Gilli\inst{\ref{inafbo}}$^{\orcidlink{0000-0001-8121-6177}}$,
K.~I.~Caputi\inst{\ref{kapteyn}}$^{\orcidlink{0000-0001-8183-1460}}$,
M.~Ricotti\inst{\ref{univmaryland}}$^{\orcidlink{0000-0003-4223-7324}}$,
C.~Gruppioni\inst{\ref{inafbo}}$^{\orcidlink{0000-0002-5836-4056}}$,
S.~Ravindranath\inst{\ref{stsci}}$^{\orcidlink{0000-0002-5269-6527}}$,
A.~Mercurio\inst{\ref{univsalerno},\ref{inafnapoli}}$^{\orcidlink{0000-0001-9261-7849}}$,
V.~Strait\inst{\ref{dawn},\ref{univcope}}$^{\orcidlink{0000-0002-6338-7295}}$,
N.~Martis\inst{\ref{nrccanada},\ref{halifax}}$^{\orcidlink{0000-0003-3243-9969}}$,
R.~Pascale\inst{\ref{inafbo}}$^{\orcidlink{0000-0002-6389-6268}}$,
G.~B.~Caminha\inst{\ref{univmunich}}$^{\orcidlink{0000-0001-6052-3274}}$,
M.~Annunziatella\inst{\ref{spain_astrob}}$^{\orcidlink{0000-0002-8053-8040}}$
}

\institute{
INAF -- OAS, Osservatorio di Astrofisica e Scienza dello Spazio di Bologna, via Gobetti 93/3, I-40129 Bologna, Italy \label{inafbo} 
\and
Dipartimento di Fisica, Università degli Studi di Milano, Via Celoria 16, I-20133 Milano, Italy\label{unimi}
\and
INAF -- Osservatorio Astronomico di Roma, Via Frascati 33, 00078 Monteporzio Catone, Rome, Italy\label{inafroma}
\and
Dipartimento di Fisica e Scienze della Terra, Università degli Studi di Ferrara, Via Saragat 1, I-44122 Ferrara, Italy\label{unife}
\and
INAF -- IASF Milano, via A. Corti 12, I-20133 Milano, Italy\label{inafiasf}
\and
University of Ljubljana, Department of Mathematics and Physics, Jadranska ulica 19, SI-1000 Ljubljana, Slovenia\label{uniLjubljana}
\and
Department of Physics and Astronomy, University of California Davis, 1 Shields Avenue, Davis, CA 95616, USA\label{unicalifornia}
\and
Department of Astronomy, Oskar Klein Centre, Stockholm University, AlbaNova University Centre, SE-106 91, Sweden\label{univstock}
\and
NSF's National Optical-Infrared Astronomy Research Laboratory, 950 N. Cherry Ave., Tucson, AZ 85719, USA\label{tucson}
\and
Max Planck Institut für Astrophysik, Karl-Schwarzschild-Straße 1, D-85748 Garching bei M\"unchen, Germany\label{maxplanck}
\and
INAF Osservatorio Astronomico di Padova, vicolo dell'Osservatorio 5, 35122 Padova, Italy\label{inafpd}
\and
NRC Herzberg, 5071 West Saanich Rd, Victoria, BC V9E 2E7, Canada\label{nrccanada}
\and
D\'epartement d’Astronomie, Université de Gen\'eve, Chemin Pegasi 51, 1290 Versoix, Switzerland\label{unigeneva}
\and
European Southern Observatory, Alonso de Córdova 3107, Casilla 19, Santiago 19001, Chile\label{esochile}
\and
European Southern Observatory, Karl-Schwarzschild-Strasse 2, D-85748 Garching bei M\"unchen, Germany\label{eso_germany}
\and
Kapteyn Astronomical Institute, University of Groningen,
P.O. Box 800, 9700AV Groningen,
The Netherlands\label{kapteyn}
\and
Department of Astronomy, University of Maryland, College Park, 20742, USA\label{univmaryland}
\and
Space Telescope Science Institute (STScI), 3700 San Martin Drive, Baltimore, MD 21218, USA\label{stsci}
\and
Dipartimento di Fisica "E.R. Caianiello," Universit\`a Degli Studi di Salerno, Via Giovanni Paolo II, I-84084 Fisciano (SA), Italy\label{univsalerno}
\and
INAF -- Osservatorio Astronomico di Capodimonte, Via Moiariello 16, I-80131 Napoli, Italy\label{inafnapoli}
\and
Cosmic Dawn Center (DAWN), Denmark\label{dawn}
\and
Niels Bohr Institute, University of Copenhagen, Jagtvej 128, DK-2200 Copenhagen N, Denmark\label{univcope}
\and
Institute for Computational Astrophysics and Department of Astronomy \& Physics, Saint Mary’s University, 923 Robie Street, Halifax, NS B3H 3C3, Canada\label{halifax}
\and
Technical University of Munich, Department of Physics, James-Franck-Stra{\ss}e 1, 85748 Garching, Germany\label{univmunich}
\and
Centro de Astrobiolog\'ia (CAB), CSIC-INTA, Ctra. de Ajalvir km 4, Torrej\`on de Ardoz, E-28850, Madrid, Spain\label{spain_astrob}
}

\date{} 

 
\abstract
{
We present \JWST/NIRSpec integral field spectroscopy (IFS) of a lensed Population III candidate stellar complex (dubbed {\it Lensed And Pristine} 1, \lap), with a lensing-corrected stellar mass $\lesssim 10^4$ \msun,  absolute luminosity M$_{\rm UV} > -11.2$ ($m_{\rm UV} > 35.6$), confirmed at redshift $6.639\pm0.004$. The system is  strongly  amplified ($\mu \gtrsim100$) by straddling a critical line of the Hubble Frontier Field galaxy cluster MACS~J0416. Despite the stellar continuum is currently not detected in the Hubble and \JWST/NIRCam and NIRISS imaging, arclet-like shapes of Lyman and Balmer lines, \lya, \hg, \hb\ and \ha\ are detected with NIRSpec IFS with signal-to-noise ratios SNR~=~$5-13$ and large equivalent widths ($>300-2000$~\AA),
along with a remarkably weak \oiiidoublam\ at SNR~$\simeq 4$.
\lap\ shows a large ionizing photon production efficiency, log($\xi_{ion}$[erg~Hz$^{-1}$])~$>$~26.
From the metallicity indexes 
R23 = (\oiii\ + \oii) / \hb~$ \lesssim 0.74$
and R3 = (\oiii\ / \hb)~$ = 0.55 \pm 0.14$,
we derive an oxygen abundance 
12+log(O/H)~$\lesssim 6.3$.
Intriguingly, the \ha\ emission is also measured in mirrored sub-components where no \oiii\ is detected, providing even more stringent upper limits on the metallicity if in-situ star formation is ongoing in this region (12+log(O/H)~$< 6$). 
The formal stellar mass limit of the sub-components would  
correspond to $\sim 10^{3}$ \msun\ or M$_{\rm UV}$ fainter than $-10$. 
Alternatively, such a metal-free pure line emitting region could be the first case of a fluorescing \hi\ gas region, induced by transverse escaping ionizing radiation from a nearby star-complex.
The presence of large equivalent-width hydrogen lines and the deficiency of metal lines in such a small region, make \lap\ the most metal poor star-forming region currently known in the reionization era and a promising site that may host isolated, pristine stars. 
}

   \maketitle

\section{Introduction}
\label{sect:intro}

\begin{figure*}
\center
 \includegraphics[width=\textwidth]{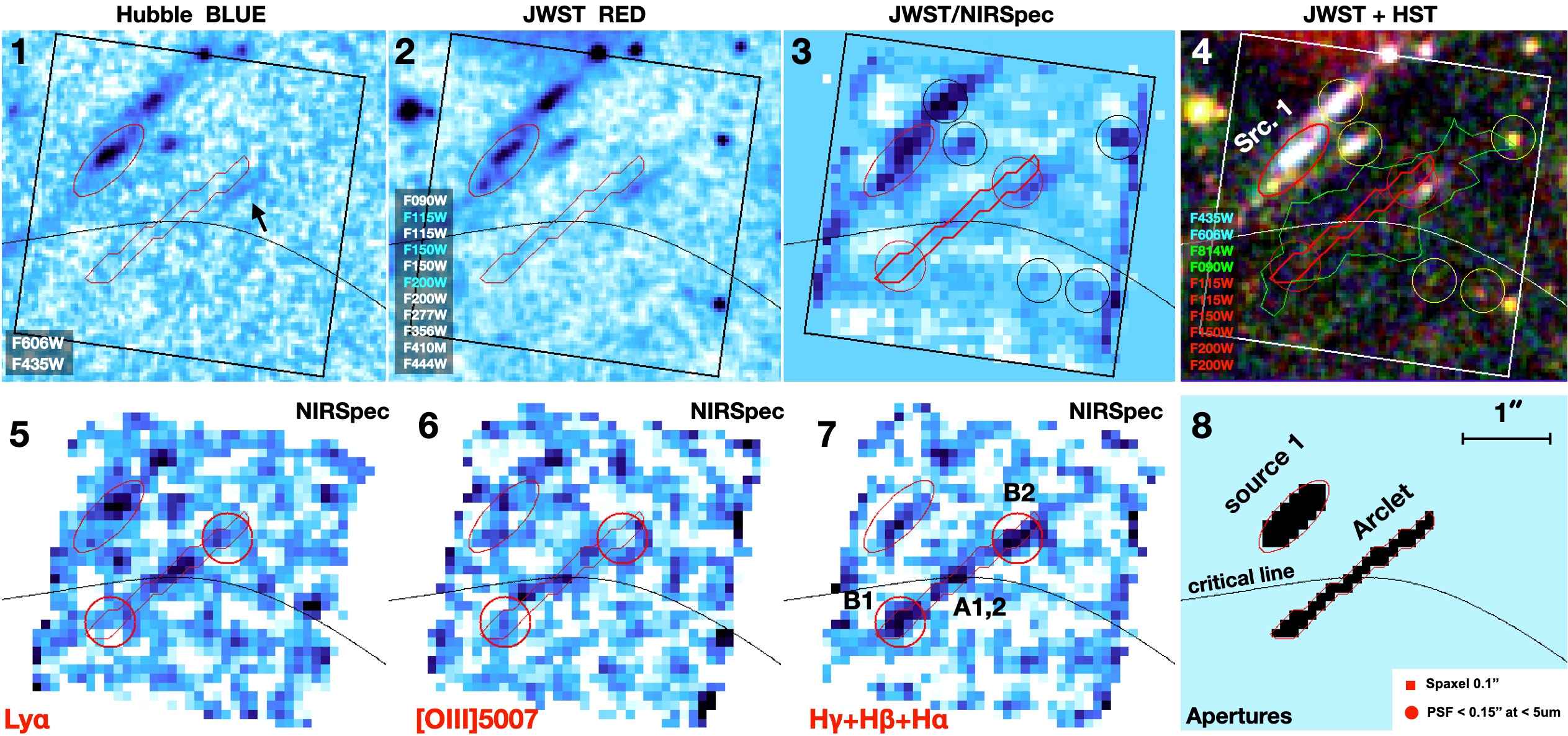}
 \caption{Photometric and spectroscopic observations of \lap. Hubble Frontier Fields \citep[][]{Lotz_2017HFF} and \JWST\ NIRISS, NIRCam imaging from the CANUCS GTO program \citep[][]{willott2022}. From top-left to bottom-right: (1) the combined HST F435W and F606W image probing $\lambda < 915$\AA\ at $z=6.639$. The black arrow indicates the presence of a foreground object which, however, is not significantly contaminating the extracted spectrum of \lap; (2) the 11-bands stacked \JWST/NIRCam and NIRISS, as labeled in the bottom left (white and cyan labeled filters indicate NIRCam and NIRISS, respectively); (3) the median collapse of the entire NIRSpec datacube spanning the range $0.7-5.2~\mu 
m$.  The black circles indicate sources visible in the \JWST\ and HST imaging; (4) the color composite image as with the three channels blue, green and red as indicated in the legend on the left (note that the red channel shows the stacked 6-bands F115W, F150W, F200W combined from NIRISS and NIRCam). The white square marks the NIRSpec FoV. The yellow circles mark the same objects as in panel (3) and the green contour outlines the $3\sigma$ 
\lya\ from VLT/MUSE. Panels (5), (6) and (7) show the \lya, \oiiiv, and the sum \hg+\hb+\ha, with the critical line (in black) and the positions of the two B1,2 components (marked with red circles), respectively; (8) the masks used to extract the spectra from the sources described 
in the text. 
The elongated red contour of $2.3'' \times 0.2''$ outlines the aperture used to extract the spectrum of the arclet (\lap), 
defined on the combined image of the Balmer lines (see Appendix~\ref{LINES}),
while the ellipse indicates the aperture used to extract the spectrum shown in Figure~\ref{Source1} and the magnitudes of Source 1 ($z=2.41$) (see Appendix~\ref{postprocessing}). All panels have the same scale, indicated in panel 8. The angular size of the single spaxel and the PSF at $5 \mu m$ is reported in the bottom-right inset of panel 8.}
 \label{fig:cutouts}
\end{figure*}

With the advent of \JWST\ the search for metal-free, population III (PopIII), sources is now entering a golden epoch. 
Although no direct observations of PopIII stars have been made to date, their existence is supported by cosmological simulations \citep[][and references therein]{abel2002, Bromm2002, hirano2014, Park2021, klessen2023} and the observation of extremely metal-poor halo stars, which are believed to be enriched by metals produced in PopIII stars 
\citep[e.g.,][]{salvadori2007, hartwig2018, vanni2023}.

Many recent papers have proposed key diagnostics aimed at identifying such elusive pristine stars. The expected presence or deficit of emission lines in PopIII sources coupled with the underlying shape of the stellar continuum affects the colors of key photometric bands, which can now be easily probed in the ultraviolet/optical rest frame up to $z\sim15$ with \JWST/NIRCAM and MIRI (e.g., see \citealt{trussler2022} and references therein). Even more informative is the direct access to ultraviolet and optical spectral features in the same early epochs (e.g., \JWST/NIRSpec and/or MIRI), derive rest-frame equivalent widths (\EW ) and key line ratios \citep[e.g.,][]{nakajima2022, nakajima2023, cameron2023, sanders2023}.
In particular, the presence of prominent helium (\EW~$>20$\AA\ rest-frame), Balmer (\EW~$>1000$\AA)
emission lines and \lya\ (if not significantly attenuated by the IGM, \EW~$>1000$\AA), along with a deficit of metal lines, support very metal-poor conditions (\citealt{nakajima2022}, see also \citealt{katz2022, Inoue2011, Zackrisson2011, schaerer2002, schaerer2003}).
However, despite the unprecedented capabilities of \JWST, identifying PopIII stars remains challenging. 
In particular, a pocket of stars forming in pristine gas conditions at early epochs (either in isolation or as a metal-unpolluted satellite of a PopII galaxy) is expected to be out of reach even for the \JWST\ sensitivity.
As a basic example, a star complex of PopIII stars at $z=7$ with a stellar mass of $10^{3,4,5}$~\msun\  corresponds an observed 1500~\AA\ ultraviolet magnitude of 37.6, 35.1, 32.6 (assuming for simplicity all stars having the same mass of 100~\msun\ and age $\lesssim 3$ Myr, \citealt{windhorst2018}).
The expected \EW~for the \heii\ line in these complexes would span the interval $20-100$~\AA\ in the rest-frame \citep[][]{POPIII_Nakajima2022}, which corresponds to fluxes lower than $5 \times 10^{-20}$ erg~s$^{-1}$~cm$^{-2}$. The strength of \heii\ may 
be significantly affected by stochastic IMF sampling, which increases the variance of the emerging line flux, or lowered by a factor 10 or more if aging, different star formation rates (SFR) or PopIII and PopII mixing are considered \citep[][]{ribas_popiii_boosting2016, katz2022, vikaeus2022}\footnote{E.g., in the fiducial model of \citealt[][]{vikaeus2022} a $10^{4}$~\msun\ system younger than 10 Myr would have a \heii\ line flux $4.5 \times 10^{-22}$ erg~s$^{-1}$~cm$^{-2}$ at $z=10.$}. These emission line flux and continuum levels 
require a significant investment of time also for \JWST\ (e.g., at $2~\mu m$ a signal-to-noise ratio of S/N = 10 for a point-like source with line flux $\simeq 4 \times 10^{-19}$ erg~s$^{-1}$~cm$^{-2}$ can be achieved with an integration time of 100,000~s, for R=1000, \citealt{jakobsen2022}).

\begin{figure*}
\center
 \includegraphics[width=\textwidth]{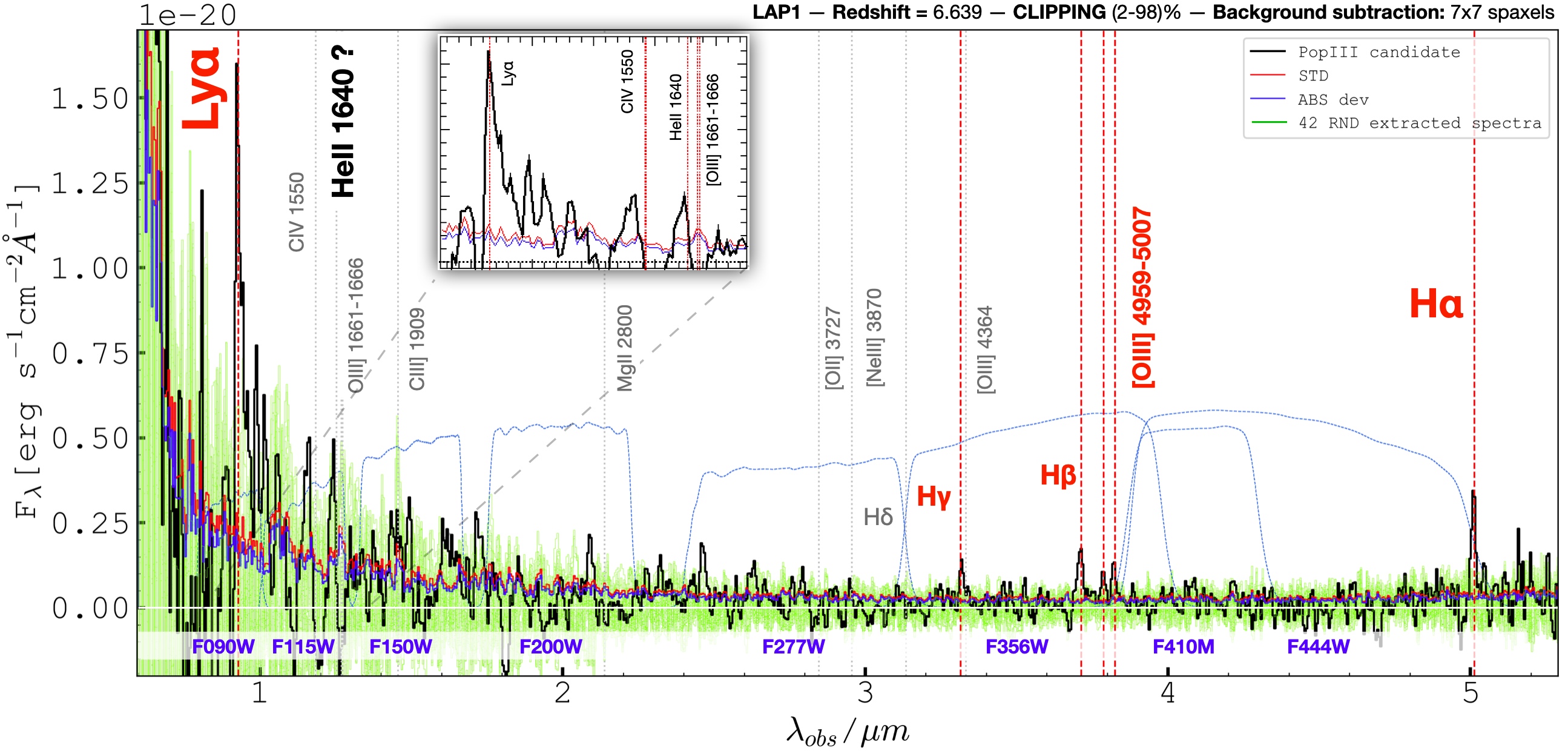}
 \caption{The one-dimensional NIRSpec spectrum of \lap\ (black line) extracted from the elongated aperture (``Arclet'') shown in panel 8 of Figure~\ref{fig:cutouts}. The most relevant lines, detected with SNR~$>4$, are indicated in bold red text and dashed lines. Additional undetected atomic transitions are shown with gray dotted lines at the redshift inferred from the Balmer lines, along with the uncertain \heii\ detection  marked in black. The green area shows the distribution of 42 spectra randomly extracted  with the same aperture within the FoV. The blue and red lines show the corresponding absolute median and standard deviations, respectively. In the top-right title, the redshift, the adopted clipping threshold, and the background window size used in the post-processing are quoted (see Appendix~\ref{postprocessing} for more details). The standard and mean absolute deviations are shown in red and blue, respectively. \JWST\ photometric bands are outlined with blue dotted lines and labeled on the bottom of the figure.}
 \label{spec_popiii}
\end{figure*}

In principle, the detection of more massive systems (e.g., $10^{5-6}$~\msun) purely composed by PopIII stars would be possible with \JWST. 
Indeed, detailed calculation at $z=8$ by \citet{trussler2022}, adopting different IMFs for a $10^{6}$ \msun\ PopIII complex produce more accessible line fluxes (\heii) and magnitudes. In the most favorable conditions (e.g., adopting their Kroupa IMF with characteristic mass of stars of 100 \msun) a few tens hours integration time with \JWST/NIRSpec or NIRCam slitless spectroscopy are needed to reach a $5\sigma$ detection of the helium line, or a few hours with NIRCam if the source is  imaged immediately after an instantaneous starburst. The exposure time diverges to hundreds of hours if the characteristic stellar mass decreases to 10 \msun\ or lower \citep[see][for more details]{trussler2022}.
Therefore, an integration time of dozens of hours would still be required, the number density of such high mass objects is still unknown and likely very low \citep[][]{klessen2023, vikaeus2022}. A mixture of PopII and PopIII components has been  
postulated (e.g., \citealt{venditti2023}) with spectral features which would reflect such a hybrid condition \citep[][]{sarmento2018, sarmento2019}. However, even though these systems would produce accessible fluxes, the signature emerging from the pockets of pristine gas in the metal-enriched galaxy would be diluted or misinterpreted, if not enough angular resolution (spatial contrast) is available. Sources showing \heii\ emission have already emerged at $z\simeq 8$ from \JWST\ first data, but further investigations are 
needed to better locate and characterize the region emitting such hard photons \citep[][see also the controversial case at $z=6.6$, dubbed CR7, \citealt{sobral2019, shibuya2018}]{wang2022}. Not 
to mention that even metal-enriched star-forming regions with SFR $< 1 $~\sfr\ can show elevated \heii\ equivalent width due to IMF sampling issues \citep[e.g.,][]{vikaeus2020} (see also \citealt{shirazi_2012_heii, senchyna20, senchyna_2021_VMS, schaerer2019_xrb, kehrig_2018, bik18} reporting on additional mechanisms/sources of \heii\ line emission, like high mass X-ray binaries, shocks or very massive stars). 

Because of these limitations, 
gravitational lensing offers a promising tool to investigate pristine stars \citep[][]{rydberg2013, zack2015, vikaeus2022, vanz_popiii}. In particular, lensing magnification increases (1) the spatial contrast and (2) the signal-to-noise ratio of the observed features, with the caveat of a limited accessible volume as the magnification increases.
Therefore, if the identification of extremely metal-poor conditions is restricted to very small spatial scales (e.g. before the targeted region is polluted by any previous or nearby/concurrent SF event), the required large spatial contrast in such studies becomes relevant. Indeed, the fact that an isolated stellar complex or cluster of extremely metal-poor (EMP) or metal-free stars is expected to be found in isolation requires special observational conditions 
\citep[][]{katz2022}.

\citet{vanz_popiii} reported the identification of a \lya-arclet at redshift $z=6.63$ straddling a critical line, with no evident detection of a stellar counterpart in deep Hubble Frontier Fields (HFF) images of the galaxy cluster MACS~J0416 \citep[][]{Lotz_2017HFF}. The \lya\ emission was detected at S/N = 17 from deep MUSE observations \citep{vanz_popiii, vanz_mdlf}, with a flux of $(4.4\pm 0.25) \times 10^{-18}$ \ergscm\ and an equivalent width likely larger than 500~\AA\ rest-frame (or 1000~\AA\ if the intergalactic medium attenuates 50\% of the line). The undetected stellar counterpart and the large amplification lead to an estimated stellar mass of $\simeq 10^4$ \msun. Though possibly rare at these redshifts (but still expected down to $z\sim 3-5$, \citealt{Tornatore2007,bromm2013, pallottini2014, liu_bromm2020}), the large equivalent widths indicate a possible presence of extremely metal-poor or even PopIII stars, making this source an ideal target for \JWST\ \citep[][]{Gardner2023_JWST, Rigby_23PASP}. In fact, at $z=6.64$ all the Balmer and the most prominent metal lines (e.g., \oiiidoublam)
can be observed in a single shot with the \JWST/NIRSpec prism observing mode \citep[][]{jakobsen2022, ferruit2022, boker23}. We present here \JWST/NIRSpec prism integral field spectroscopy observations of such an exotic source, covering the rest-frame spectral range from \lya\ to \ha, in addition to \JWST/(NIRCam + NIRISS) imaging.

Throughout this paper, we assume a flat cosmology with $\Omega_{M}$= 0.3,
$\Omega_{\Lambda}$= 0.7 and $H_{0} = 70\,{\rm km}\,{\rm s}^{-1}\,{\rm Mpc}^{-1}$. 
All magnitudes are given in the AB system \citep{Oke_1983}:
$m_{\rm AB} = 23.9 - 2.5 \log(f_\nu / \mu{\rm Jy})$.


\section{\JWST\ and \HST\ imaging: still an undetected source} \label{sec:imaging}

\JWST/NIRCam and \JWST/NIRISS observations were acquired on January 2023,  as part of the 
CAnadian NIRISS Unbiased Cluster Survey: CANUCS (\citealt[][]{willott2022}). The galaxy cluster MACS~J0416 was observed in eight NIRCam filters covering the spectral range from $0.8\mu m$ to $5\mu m$ (F090W, F115W, F150W, F200W, F277W, F356W, F410M, F444W) and with an integration time of 6400 seconds per band. 
\JWST/NIRISS imaging was also acquired as part of the pre-imaging for slitless spectroscopy in the F115W, F150W and F200W bands for an integration time of 2280 seconds per band. The images were processed using a combination of the STScI JWST pipeline v1.8.4 with CRDS context jwst\_1027.pmap and \textsc{GRIZLI} v1.7.8 (\citealt[][]{Grizli}). A more detailed description of the CANUCS imaging processing will be presented in Martis et al. (in prep.).  The magnitude limits for point sources at 5-$\sigma$ are 29.4 and 29.1 in the F150W band for \JWST/NIRCam and \JWST/NIRISS, respectively. Hubble Frontier Fields data in the F435W, F606W, F814W, F105W, F125W, F140W, and F160W bands are also included in the set of images used in this work, all of them with a typical 5-sigma magnitude limit for point sources of $\simeq 29$ \citep[][]{Lotz_2017HFF}.

\begin{figure*}
\center
 \includegraphics[width=17cm]{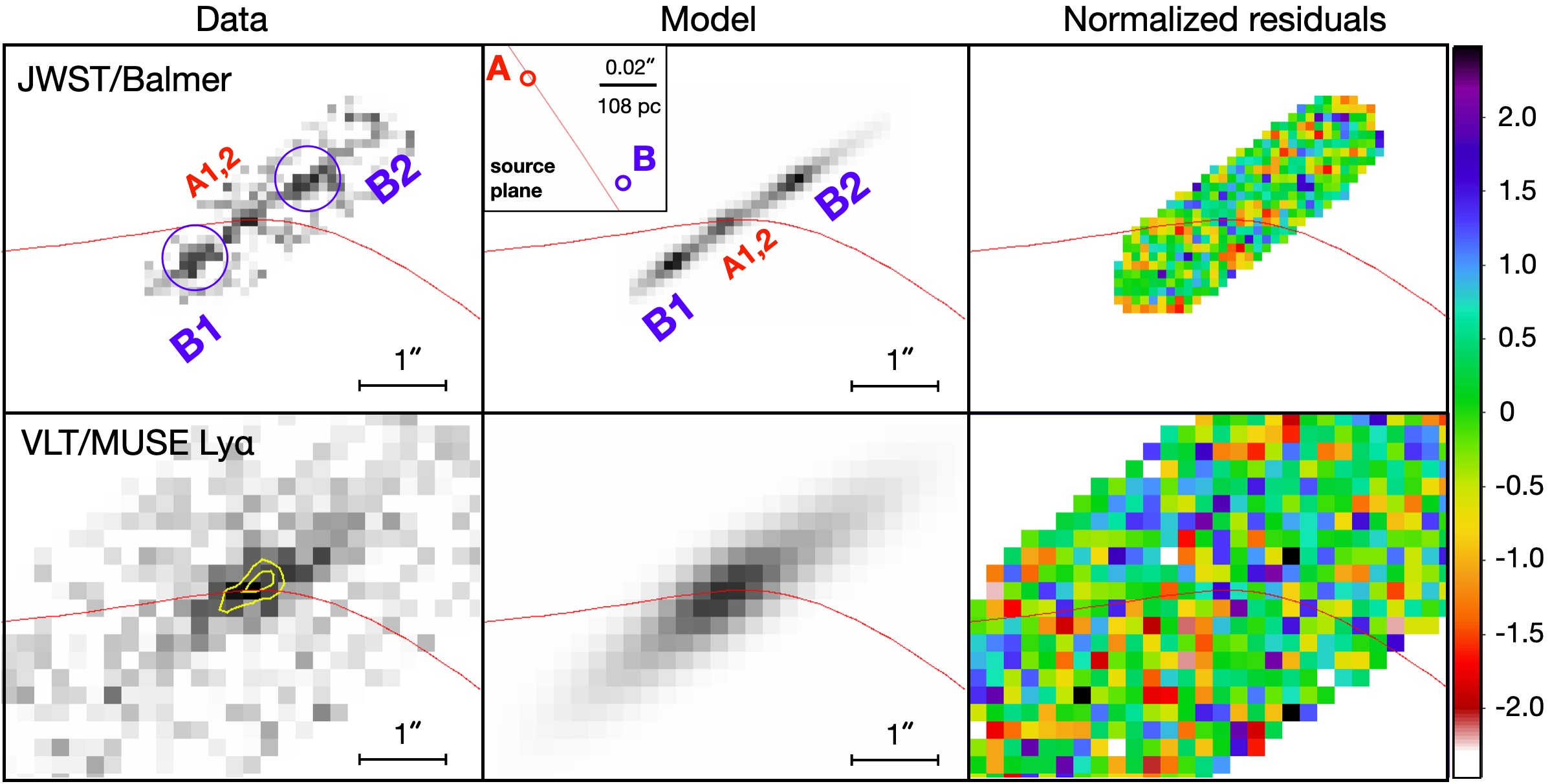} 
 \caption{Results from the modeling of the lensed arclet \lap\ using our forward modeling procedure. The top-left panel shows the co-added flux of the Balmer lines (\hg+\hb+\ha) from the NIRSpec IFS datacube, where the blue circles indicate the position of the B1,2 mirror images and the red curve is the critical line corresponding to a source redshift $z=6.64$ passing through the merged A1,2 images. The top-mid panel shows the corresponding model image; the inset shows the source plane configuration of A and B, where we also indicate the physical scale and the caustic line. The top-right panel shows the normalized residuals, ${\rm (Data-Model)}/\sigma$. The bottom panels similarly show the {\tt GravityFM} modeling of the VLT/MUSE \lya\ image, with overlaid \lya\  emission contours (in yellow) obtained from the NIRSpec datacube.} 
 \label{fig:FM}
\end{figure*}

Figure~\ref{fig:cutouts} shows the stacked images centered at the coordinates of the \lya\ arclet, which is hereafter referred  to as \lap\ (Lensed And Pristine 1).
Remarkably, and similarly to what is shown by \citet{vanz_popiii}, there is no evidence of stellar counterparts in the proximity of the arclet, in any set of images, HST, \JWST/NIRcam or \JWST/NIRISS, neither in the deep stacked image which collects F115W, F150W, and F200W (after combining NIRCam and NIRISS) probing the $2000$~\AA\ rest-frame, nor in the redder bands ($\lambda >2\mu m$), F277W, F356W, F410M and F444W probing the optical 4000\AA\ 
rest-frame. The deep stacked ultraviolet and optical images provide lower limits of m$_{\rm UV}$ and m$_{\rm opt}$ $\simeq 30.4$ at $2\sigma$ ($\simeq 31$ at $1\sigma$), when measuring the flux with the A-PHOT tool \citep[][]{merlin19} in an elliptical aperture which includes the full arclet shape. The magnitude limits are reported in Table~\ref{tab:spec}
 and are used in Sect.~\ref{sec:discussion} to constrain the limits on the equivalent widths of the emission lines and the intrinsic luminosity. 
The de-lensed properties of {\em LAP1} are reported in Sect.~\ref{FM}.

Remarkably, despite the absence of any stellar counterpart, the \JWST/NIRSpec observations reveal four emission lines, \lya, \hg, \hb, \oiiidoublam\ and \ha, which follow the arclet-like shape already outlined by the previous VLT/MUSE \lya\ emission at $z=6.64$.

\section{JWST NIRSpec/IFU observations}

\JWST/NIRSpec integral field unit (IFU) observations were performed on October $16-17$, 2022, as part of five pointings targeting strongly lensed globular cluster precursors and candidate population III stellar complexes at redshift $z=6-7$ (PI Vanzella, prog. id 1908). In particular, four out of five pointings will cover a structure of tiny star-forming regions and proto-globulars at $z=6.14$ \citep[][]{vanz19,vanz_mdlf,calura2021}, for a total integration time of $\simeq 18$h (currently scheduled for Summer 2023). 
One out of five pointings was
devoted to \lap\ at $z=6.64$. Here we present observations on \lap, which focus on an extremely faint candidate PopIII star complex discovered by \citet{vanz_popiii}. A total integration time of 6.19 hours (including overheads) split into eight independent acquisitions of 2115.4 seconds each were performed, for a net integration time on target of 4.7 hours. The small ($0.25"$ extent) dithering cycling over 8 points was applied to each acquisition.

 \begin{figure}
\center
 \includegraphics[width=\columnwidth]{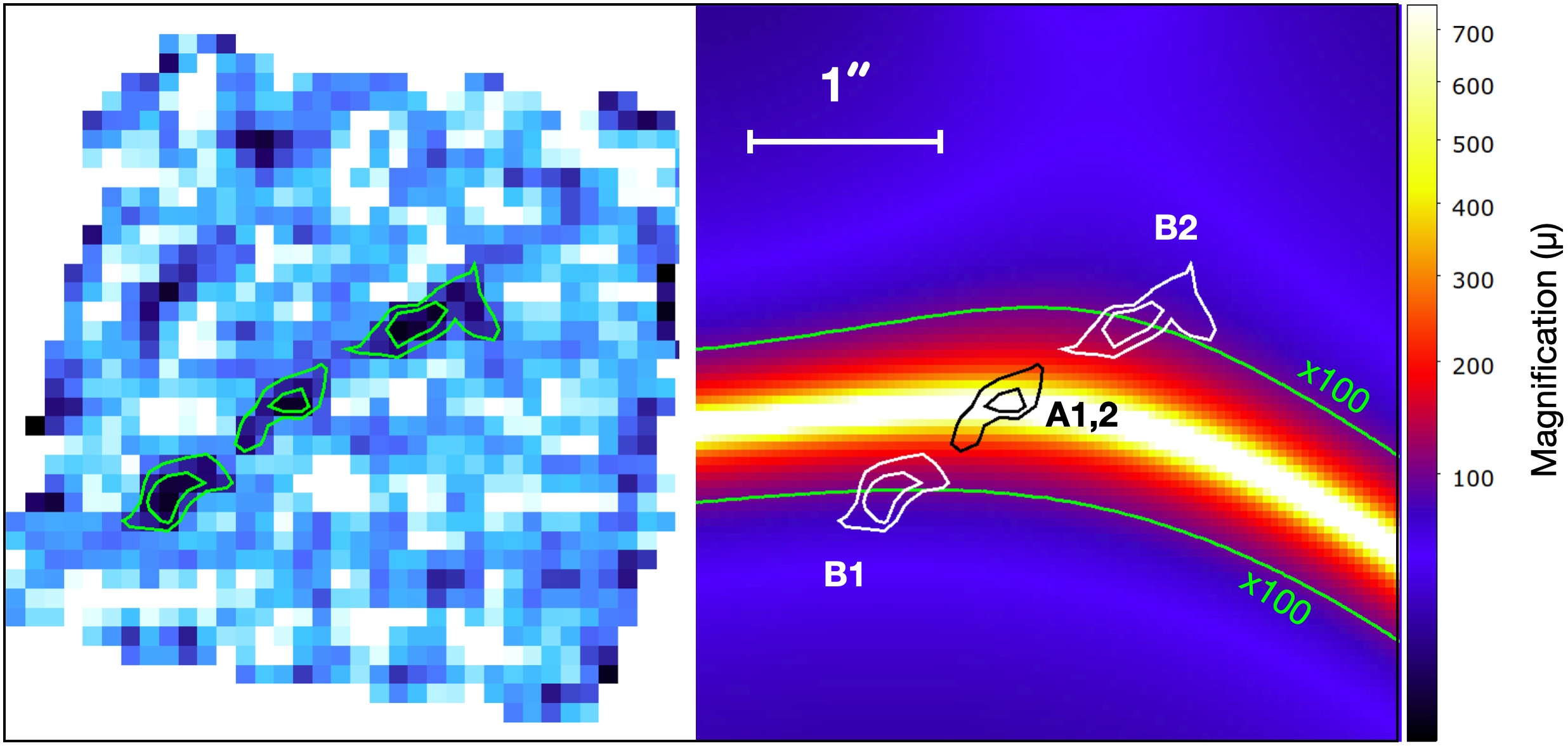} 
 \caption{The magnification map extracted from the $\mathrm{B23_{new}}$ lens model is shown on the right panel, with contours marking the stacked triply imaged Balmer line emission produced by the A and B components (see the left panel with the same contours). The two mirrored images B1 and B2, and A1,2 are labeled on the color coded $\mu_{tot}$ map (adopting a square root scale). The lines corresponding to $\mu_{tot} = 100$ are also overlaid in green.} 
 \label{fig:mumap}
\end{figure}

\subsection{Data Reduction}
\label{sect:reduction}

We reduced the NIRSpec/IFU raw data using the STScI pipeline (version $1.9.5$). The software version and the Calibration Reference Data System (CRDS) context are $11.16.20$ and jwst\_1062.pmap, respectively. 
The raw data (i.e., \textit{uncal} exposures) were processed through three stages. To summarize, Stage 1 is common to all the JWST instruments and corrects for detector--related issues (e.g., bias and dark subtraction, pixel saturation and deviation from linearity, cosmic rays flagging). Stage 2 converts the coordinates from the detector plane to sky coordinates and implements critical steps such as background subtraction, flat--field correction\footnote{At the time of writing, this step can introduce a systematic uncertainty as high as 10\% due to a simplification in the reference files \citep{boker23}.}, flux and wavelength calibration. Also, intermediate datacubes associated with the single dithers are produced in this stage. The third and last stage combines the eight calibrated dithers into the final datacube.

We ran the three stages using the default parameters. We carefully inspected the intermediate products for each step. In particular, we did not find in the count-rate images (i.e., \textit{rate} files) produced after the first stage any significant vertical pattern associated with correlated noise (e.g., \textit{1/f} noise), as discussed by other works using NIRSpec/IFU (e.g., \citealt{marshall23, ubler23}), and thus we did not apply any correction outside the pipeline before running Stage 2. Moreover, we did not perform any background subtraction inside the pipeline, but used an independent procedure as described in Appendix~\ref{postprocessing}.
Finally, we ran the \textit{Outlier Detection Step} in Stage 3, which is meant to remove possible cosmic rays not recognized in the \textit{Jump Step} of Stage 1, by comparing the single exposures. Other studies have shown that this step can lead to false positive cosmic rays corresponding to bright sources in dithered exposures  (i.e., quasars, see \citealt{cresci23, marshall23, perna23}), thus skipping this step is often recommended (see also \citealt{boker23}).  
However, by comparing the products obtained by applying and skipping the \textit{Outlier Detection} procedure, we found that spurious cosmic rays are not produced at the location of the sources, likely due to the low flux regime of our targets. We verified that the application of this step with the default parameters works as expected by removing several spikes affecting our dataset. The final calibrated datacube is thus obtained after completing the three-stage data processing. We used the default spatial scale of $0.1"$ for each spaxel.  

Additional post-processing was performed on the reduced datacube. This includes background subtraction, the removal of outliers, and the computation of the error spectrum. In addition, cross-checks on the flux calibration were performed using \JWST\ NIRCam, NIRISS and Hubble photometry on sources lying in the same field of view (see Appendix~\ref{postprocessing} for more details).
Finally, the detected sources in the reduced, post-processed and collapsed data-cube have been aligned to the \JWST/NIRCam counterparts by applying a rigid shift on RA and DEC (see the sources marked with circles in Figure~\ref{fig:cutouts}).

%
\begin{table}
\caption{Observed and derived properties of \lap.} 
\label{tab:spec}      
\centering          
\begin{tabular}{l c c c}     
\hline\hline  
Source    &   Parameter  &  Value   &     \\ 
\hline
\lap &\ha & $69.4 \pm   5.5 $ & $10^{-20} cgs$ \\ 
&\oiiiv  & $  14.5 \pm   3.4 $ & $10^{-20} cgs$  \\ 
&\oiiiiv & $  9.8 \pm   3.2 $ & $10^{-20} cgs$ \\ 
&$(\dag)$~\oiiiiv$^{\star}$ &  4.9 & $10^{-20} cgs$  \\ 
&\hb & $  26.3 \pm   2.7 $ & $10^{-20} cgs$ \\ 
&\hg & $  12.8 \pm   2.4 $ & $10^{-20} cgs$  \\ 
&\oiilam & $  <  3.0 $ & $10^{-20} cgs$ \\ 
&$(\ddag)$~\heii & $  79.6 \pm  20.7 $ & $10^{-20} cgs$ \\ 
&\lya  & $ 369.2 \pm  29.3 $ & $10^{-20} cgs$\\ 
&\EW(\lya) & $> 370$ & \AA, $3\sigma$, rest  \\
&\EW(\ha) & $> 2020$ & \AA, $3\sigma$, rest \\
&\EW(\hb) & $> 420$ & \AA, $3\sigma$, rest  \\
&\EW(\oiiiv) & $> 246$  & \AA, $3\sigma$, rest \\ 
\hline
& \ha/\hb & $   2.64_{-  0.31}^{+  0.37}$ & $-$  \\ 
& \hb/\hg & $   2.05_{-  0.37}^{+  0.53}$ & $-$ \\ 
& \oiiiv/\hb & $   0.55_{-0.14}^{+0.15}$ & $-$ \\ 
& \lya/\ha & $   5.32_{-  0.56}^{+  0.63}$ & $-$  \\ 
& R23 &    $0.74_{-  0.19}^{+  0.21}$ & $-$ \\ 
& R3  &    $0.55_{-  0.13}^{+  0.14}$ & $-$ \\ 
& O32 & $  > 3(6)$ & 2(1)$\sigma$ \\ 
 \hline
&m$_{\rm UV}$[2000\AA]  & $> 30.4$ & $2\sigma$  \\
&m$_{\rm opt}$[4000\AA] & $> 30.4$ & $2\sigma$  \\
\hline
&Z &  $ <0.004$ & $Z_{\odot}$  \\ 
&12+log(O/H) &$<6.3$  & O32~$>6$ \\ 
&log($\xi_{ion}$)  & $>26$ & [erg Hz$^{-1}$], $2\sigma$\\
\hline
&$\mu_{tot}$ & $120^{+9}_{-9}$ &  median  \\
&$\mu_{tang}$ & $55^{+2}_{-6}$ & median \\
\hline \hline
A & $\mu_{tot}$ & $> 500$ & A1,2 \\  
B &$\mu_{tot}$ &  $98,99$ & B1,B2 \\ 
B1+B2 & R3 & $<0.2$ & \ha-based\\
B1+B2 & R23 & $<0.4$ &\ha-based\\
B1,2    &m$_{\rm UV}$[2000\AA] & $>32.4$ & (B1 or B2),$2\sigma$  \\
\hline \hline
\end{tabular}
\tablefoot{The reported errors are at $1\sigma$ confidence level, if not specified. The observed \lya\ line is likely attenuated by the IGM. The  values indicated with units {\it cgs} refer to fluxes with units erg~s$^{-1}$~cm$^{-2}$.
($\dag$)~This is the flux of \oiiiiv\ inferred from the observed \oiiiv\ by adopting the intrinsic flux ratio (\oiiiv/\oiiiiv)=2.98 \citep{StoreyZeippen2000}, and used to derive the O32 and R23 indexes described in the text. $(\ddag)$~The derived  flux is very tentative (see text for details). In the main text we consider the $1\sigma$ upper limit of the \heii\ line flux as indicated in column ``Value'' ($1\sigma,~20.7 \times 10^{-20}$~cgs). The intrinsic fluxes and magnitudes can be derived by dividing the reported observed values by $\mu_{tot}$. The coordinates of \lap\ are RA~$=64.0457716$, DEC~$=-24.0601283$.
}
\end{table}

\subsection{Prominence of Balmer and deficit of metals lines}

As discussed in \citet{vanz_popiii}, the arclet was detected only in \lya\ emission, without showing any significant stellar-continuum counterparts. The faintness of such an object is confirmed also by the \JWST\ imaging as reported in Sect.~\ref{sec:imaging}. However, the \JWST\  spectroscopic data provide a wealth of unique information which is key to the physical interpretation of the source. 

The \ha, \hb\ and \hg\ emission lines emerge at the mean redshift $z=6.639$ with a standard deviation of 0.004, resembling the same arclet-like orientation initially reported by \citet{vanz_popiii}.\footnote{The inferred redshift appears slightly higher than the value obtained from VLT/MUSE, z=6.629. This discrepancy is not fully clear at the time of writing. We note, however, that the \JWST/NIRSpec wavelength calibration is still under verification  especially in the case of the IFS mode \citep[see Sect.~6.3 in][]{boker23}. This possible shift does not affect the conclusions of this work.} Besides \lya, also the Balmer lines \hg, \hb\ and \ha\ are detected, along with a remarkably faint \oiiidoublam. 
The arclet is considerably thin and appears not resolved along the radial direction.\footnote{It is worth noting that NIRSpec spaxel angular scale of $0.1''$ undersamples the \JWST\ PSF at $\lambda < 2.5 \mu m$.} 
{An elongated aperture has been defined using the stacked two-dimensional images of the emerging Balmer lines. In particular, the aperture extends $2.3''$ along the arclet (tangential direction) and $\simeq 0.2''$ perpendicularly (radial direction), as outlined in Figure~\ref{fig:cutouts} and nearly marks the $3\sigma$ contour of the arclet in the stacked image.

Figure~\ref{spec_popiii} shows the one-dimensional spectrum of \lap\ and the line fluxes and errors are reported in Table~\ref{tab:spec}. Hydrogen lines \lya, \hg, \hb\ and \ha\ are detected with SNR spanning the range $5-13$ with unconstrained velocity widths, being all of them consistent with the prism spectral resolution. 
The low spectral resolution provided by the prism mode $R=30(300)$ at $\lambda =1(5)~\mu m$ corresponds to $dv \simeq 10000(1000)$~\kms~and prevents us from measuring a velocity offset among the spectral features.
The line ratios  \ha/\hb\ and \hb/\hg\ are consistent within the uncertainties with the case B recombination, corresponding to \ha/\hb~$\simeq 2.8$ and \hb/\hg~$\simeq 2.1$, \citep[e.g.,][]{osterbrock1989}, implying very little or no dust attenuation. Under the same case B assumption, the expected \lya/\ha\ is 8.7, suggesting that nearly half of the \lya\ line is attenuated by the IGM (after neglecting any additional internal absorption or geometrical effect). Remarkably, the extracted spectrum from \lap\ shows an \oiiiv\ emission significantly fainter than \hb, with \oiiiv / \hb\ $\simeq 0.55$. 
It is worth noting that such a deficit of oxygen emission compared to the Balmer lines is opposite to the strong optical oxygen emission recently observed at $z>5-6$ with \JWST, in which large  equivalent widths of \oiii\ ($>1000$\AA\ rest-frame) showing \oiiiv / \hb~$\gg1$ are commonly observed \citep[e.g.,][]{matthee22_nircam_slitless} (see also \citealt{withers2023_canucs, rinaldi2023, williams2023, endsley21, boyett_2022a, castellano17, barros2019}).

Figure~\ref{spec_popiii} also shows a possible detection of the \heii\ transition, which is a key feature related to the possible hardness of the underlying spectrum, 
expected for PopIII ionizing sources (see Sect.~\ref{sect:intro}). 
Although the feature has a formal peak SNR~$\sim3.8$ (see the inset in Figure~\ref{spec_popiii}), we notice a small blueshift relative to the Balmer lines, $dv \simeq -1200$~\kms. Such a shift corresponds to only $\sim$1/4 of the native NIRSpec prism dispersion at 1.253$\,\mu$m, however a real velocity offset might weaken the reliability of its identification as \heii. 
Moreover, the measured flux (reported in Table~\ref{tab:spec}) would imply a rest-frame equivalent width $\gtrsim 200$ \AA\ and $\log{\rm (\heii/\ha)} = 0.06$. Such values would be quite extreme even for a PopIII scenario, lying at the limits of the proposed ranges \citep[e.g.,][]{katz2022, POPIII_Nakajima2022}. We adopt a more conservative approach by considering that the \heii\ line is currently not detected. The \heii\ and continuum non-detections make the equivalent width unconstrained, while the comparison with \ha\ gives $\log{\rm (\heii/\ha)}\!<\! -0.5$. However, the different geometry among the emitting regions might affect this ratio and will require a deeper study.
A non-detection of \heii, however, does not exclude the PopIII scenario.\footnote{Note that the other helium line \heiiopt\ is expected to be 8 times fainter than \heii\ and it is unconstrained by the current observations.}

Another emission feature with similar SNR peaks approximately 200\AA\ blueward of the expected wavelength of \civ, and we similarly conclude that it is likely to be spurious.

Before discussing the implications of the physical properties of \lap\ from the line ratios and magnitude limits listed in Table~\ref{tab:spec}, a careful de-lensing is required to clarify what portion of the star-forming region we are observing.

\section{Source-plane Reconstruction}
\label{FM}

\lap\ is an arclet confirmed at $z=6.639$ detected by means of nebular lines, without a significant detection of a stellar continuum counterpart. To infer the intrinsic properties of the source, we use the lens model recently presented in \citet{bergamini_2022_J0416}. 
The model suggests that the region where \lap\ lies has a multiplicity higher than one. As a result, we expect multiple images at the redshift and location of the source, with a third less magnified counter-image (more than 10 times fainter) falling on the opposite side of the galaxy cluster. 
A high multiplicity in the region surrounding the arclet (within $6''$) is also corroborated by the presence of multiple images confirmed with deep VLT/MUSE observations at slightly lower redshift, $z=6.14$ \citep[][]{vanz_paving,vanz19}. 
In \citet{vanz_popiii} we found that the \lya\ emission of \lap\  observed with deep VLT/MUSE (seeing-limited with a FWHM of $0.6''$) is well reproduced by considering an emitting object on the source plane, very close 
to the tangential caustic, which generates two mirrored images straddling the critical line on the lens plane. These two images are too close to be individually resolved by MUSE, thus resulting in the observed single \lya\ arclet. 

The \JWST/NIRSpec observations add interesting details on the shape and structure of the observed arclet. As shown in Figure~\ref{fig:cutouts} (and Appendix~\ref{LINES}), a triple knot morphology aligned along the tangential stretch emerges when the Balmer lines are stacked all together.
Specifically, since we expect two multiple images at the location of the arclet, the presence of three knots where the critical line is expected suggests that there are two source components on the source plane: one (dubbed B), which generates the two knots at the edges of the arclet (B1 and B2), and another one (named A), superimposed to the caustic, producing two merged images on the critical line, appearing as a single knot on the lens plane and not resolved by \JWST/NIRSpec (see Figure~\ref{fig:cutouts}). Thus, the critical line likely passes through the central knot of \lap.
The observed configuration is accurately modeled in the next section. 

\subsection{Forward modeling with {\tt GravityFM}}
The proximity of the source to the critical line challenges any lens-model prediction and makes direct source reconstruction difficult, since the magnification gradients around the source location are very large. 
The accurate determination of the de-lensed physical properties of the background sources producing the observed morphology of the arc crucially depends on the robustness of the cluster lens model. For this reason, in our analysis we make use of the latest lens model of MACS J0416 \citep{bergamini_2022_J0416}, which is constrained by 237 spectroscopically confirmed multiple images (from 88 background galaxies), the largest dataset used to date for the total mass reconstruction of a galaxy cluster. This model is also characterized by a very small scatter between the model-predicted and observed positions of the multiple images:  $\Delta_{\rm rms}^{TOT}=0.43\arcsec$. We note that this model originally included two point-like images at the position of the arc, with large associated positional errors, based on seeing-limited MUSE data. To further improve the model accuracy in the vicinity of \lap, and specifically the position of the critical line 
\citep[see][]{vanz_popiii}, we re-optimized the \citet{bergamini_2022_J0416} model by replacing the previously adopted two images with the mirror images B1 and B2, using a small positional error based on the \JWST\ Balmer line image (see Figure~\ref{fig:FM}). This new model (hereafter $\mathrm{B23_{new}}$) still preserves a total $\Delta_{\rm rms}^{TOT}=0.52\arcsec$, with the positions of B1 and B2 very accurately predicted, i.e. with a $\Delta_{\rm rms}=0.04\arcsec$. The $\mathrm{B23_{new}}$ model is used for the forward modeling of \lap\ and to derive the magnification maps in the region of the arc.  

We use the forward-modeling tool \texttt{GravityFM} to fit the surface brightness distribution of \lap\ obtained by co-adding the Balmer lines flux within the NIRSpec IFU. This novel method, based on the python library \texttt{pyLensLib}~\citep{2021LNP...956.....M} and already used in  \citet{bergamini_2022_J0416},  will be described in detail in a forthcoming paper by Bergamini et al. (in preparation). In short, by implementing a Bayesian approach and using the deflection maps from a cluster lens model, \texttt{GravityFM} reconstructs the structural parameter and associated errors of one or more background sources by minimizing the residuals between their extended model-predicted and observed multiple images. With \texttt{GravityFM}, one can add local corrections to the deflection maps of the cluster lens model. However, in our analysis, we assume the deflection field to be fixed, a decision motivated by the aforementioned high precision of the $\mathrm{B23_{new}}$ macro model in reproducing the positions of all multiple images, including B1 and B2.
Figure~\ref{fig:FM} shows the results of this forward model optimization.  The triple knot morphology is well reproduced with two circular components parameterized as S\'ersic profiles on the source plane. The first component, 
named A, produces two unresolved images falling on the critical line (A1,2 in Figure~\ref{fig:FM}), while the second component, B, is responsible for the two mirrored  emissions B1,2 on both sides of the critical line, each of them marginally elongated along the tangential direction (note that the telescope PSF at $5\mu m$ is of the order of 1.5 spaxel, see Figure~\ref{fig:FM}). In the optimization procedure, position, total luminosity, effective radius, and S\'ersic index of both sources are free parameters, which are left free to vary with large uniform priors. 

The residuals in Figure~\ref{fig:FM}, normalized by the noise measured on the image plane, show the goodness of the fit of the extended source. The bottom panel shows that the same model, optimized for the Balmer lines, well reproduces also the MUSE \lya\ emission, in which the single multiple images are clearly not resolved. On the other hand, the NIRSpec \lya\ emission (see overlaid yellow contours) indicates that the \lya\ region is close to the central knot of \lap.
Although the detailed geometrical configuration on the source plane is still uncertain and based on emitting lines only, it is worth noting that the individual components A and B are separated by $\sim 150$ parsec on the source plane, with a formal statistical error smaller than 10\%\, and have estimated sizes of $< 5$ pc and a few tens of parsecs, respectively. 

Figure~\ref{fig:mumap} shows the magnification map on \lap, with overlaid the contours of the observed components A1,2 and B1,2. The resulting best magnification factors at the median model-predicted positions of B1 and B2 are $98_{-4}^{+5}$ and $99_{-6}^{+6}$, respectively, with relative statistical errors of $\sim\!10\%$. The 68\% central intervals have been derived by extracting the total(tangential) magnification $\mu_{tot}$($\mu_{tang}$) at the model predicted positions over 500 realizations of the lens model, by sampling the posterior probability distribution function with a Bayesian Markov chain Monte Carlo (MCMC) technique. Such statistical errors do not include systematic errors, which likely dominate the uncertainty in this large amplification regime \citep[e.g.,][]{Meneghetti_2017}. 
Similarly, from the same 500 
magnification maps, the best magnification of \lap\ calculated within the elongated aperture (shown in Figure~\ref{fig:cutouts}) is $\mu_{tot} = 120 \pm 9$.
This value is used when delensing the physical properties of \lap. The corresponding median tangential stretch is also large, $\mu_{tang} \simeq 55$  (see Table~\ref{tab:spec}). The magnification of component A is formally very large, $\mu_{tot}(A1,2)>500$. In the next section, we discuss the physical properties of the full arclet \lap, and report on its sub-components, A and B, in Sect.~\ref{popiii_complex}.

\begin{figure}
\center
 \includegraphics[width=8.4cm]{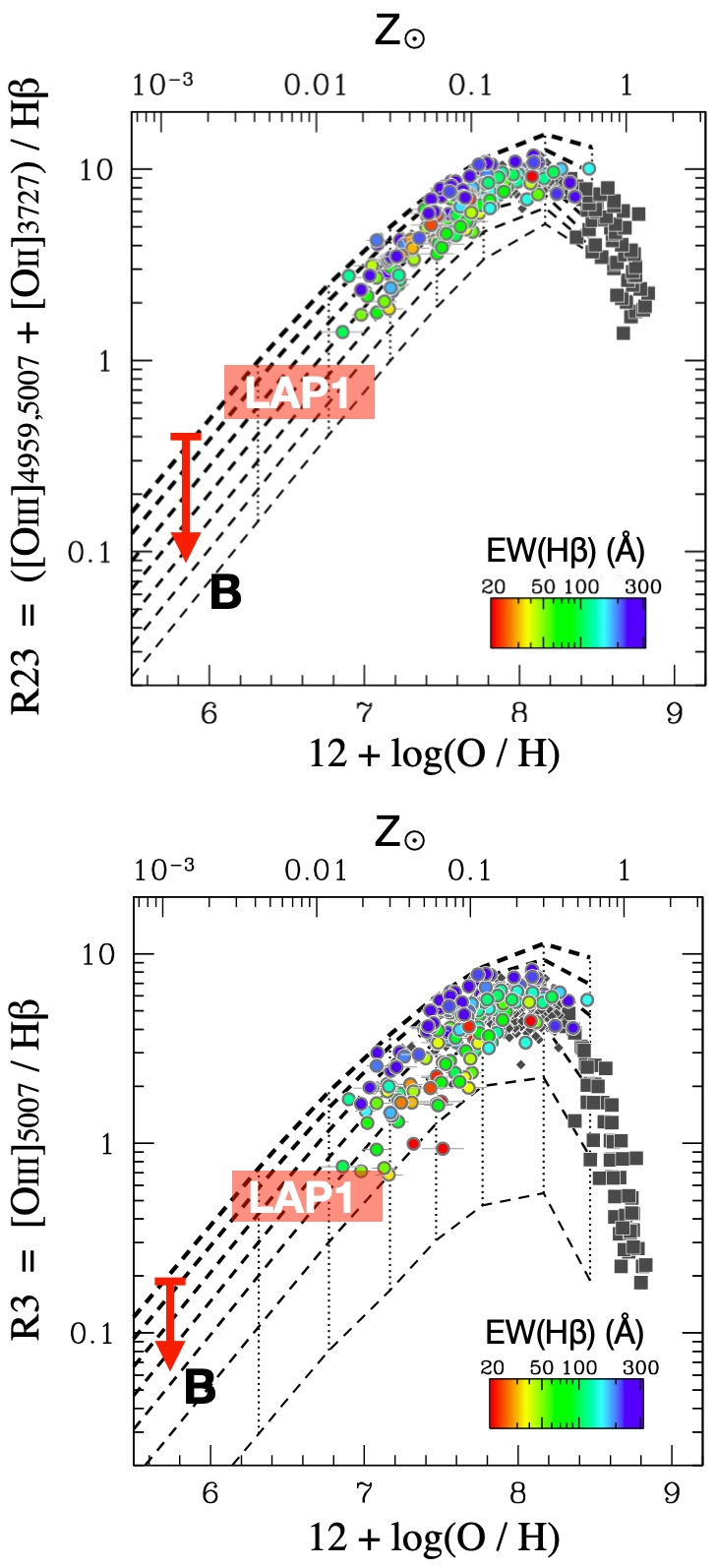}
 \caption{Photoionization model predictions of the R23 (top) and R3 (bottom) indexes based on the binary evolution SEDs (BPASS). The model tracks (dashed lines) span the ionization parameter log(U) from $-$3.5 (thin lines) to $-$0.5 (thick lines) with a step of 0.5 dex.
 The gas density is fixed to 100~cm$^{-3}$. The models encompass the scatter of data points and the dependency on the EW(\hb) by changing the ionization parameter. The datapoints are those collected by \citet{nakajima2022} for which accurate metallicities were measured with the direct $T_e$ method (figures adapted from \citealt{nakajima2022}).}
 \label{fig:R23}
\end{figure}

\section{Discussion}
\label{sec:discussion}

It is worth stressing that the source plane reconstruction is based only on the nebular emission lines, as no stellar counterpart has been detected yet. The two main emitting regions (A and B), which are separated by only $\sim 150$ parsec (see source plane diagram in Figure~\ref{fig:FM}), can be part of a larger complex or isolated. However, regardless of its morphological structure, the magnified region
shows a remarkable deficit of oxygen lines compared to the Balmer emission (e.g., \oiiiv / \hb~$ \simeq 0.55$, see Table~\ref{tab:spec}), implying a very low gas-phase metallicity.
It is also worth emphasizing that such a result is lens-model-independent, since it is based on flux ratios and is not sensitive to possible flux calibration issues (which appear minimal anyway, see Appendix~\ref{postprocessing}) given the small wavelength differences between the Balmer and metal lines.

\begin{figure*}
\center
 \includegraphics[width=\textwidth]{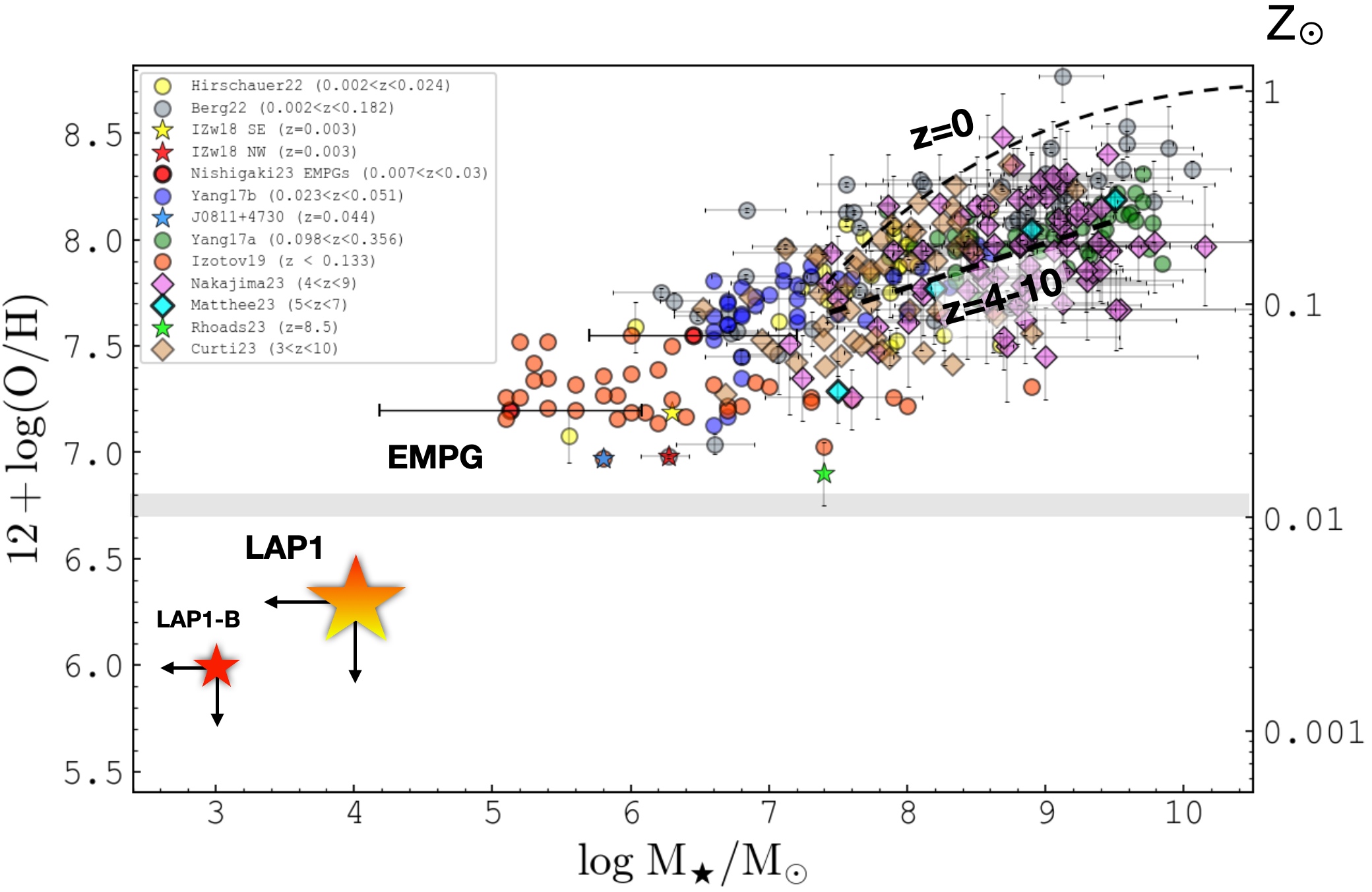}
 \caption{Gas metallicity vs. stellar mass from a collection of different surveys and categories of sources, in the local and high redshift Universe \citep[as reported in the legend,][]{Hirschauer2022, Berg2022, Izotov2018, nishigaki2023, Yang2017blue, Yang2017green, Izotov2019, nakajima2023, matthee22_nircam_slitless, Rhoads2023, curti2023}. The star-forming region described in this work, \lap\ at $z=6.639$ and its sub-component B, are marked with stars on the bottom-left and lie below the extremely metal-poor plateau identified to date (EMPG), roughly indicated with an horizontal shaded stripe.}
 \label{fig:MZ}
\end{figure*}

\subsection{A low-luminosity and low mass efficient ionizing emitter}

The $2\sigma$ magnitude limit of the arclet (\lap) inferred from the stacked \JWST\ images of 30.4 (Sect.~\ref{sec:imaging}) and the median magnification $\mu_{tot}\simeq120$ extracted from the same region (Sect.~\ref{FM}) correspond to a rest-frame de-lensed ultraviolet magnitude at 2000~\AA\  $m_{\rm UV}> 35.6$, or an absolute M$_{\rm UV}$ fainter than $-$11.2. 
%
%
This luminosity corresponds to a stellar mass of $5\times 10^{3}$~\msun, if we adopt a {\tt Starburst99} \citep[][]{leitherer14}
instantaneous burst scenario, a Salpeter IMF, and the lowest available metallicity (1/20 $Z_{\odot}$) with an age younger than 10 Myr.\footnote{This value was inferred after rescaling the {\tt Starburst99} reference case of a 10$^{6}$\msun\ system (which produces a peak 1500\AA\ luminosity of L1500~$\simeq 10^{39.5}$~erg~s$^{-1}$~\AA$^{-1}$) to the  L1500~$<~10^{37.2}$~erg~s$^{-1}$~\AA$^{-1}$ 
of \lap. The dust extinction is neglected here and the magnitude correction {\it dm} from 2000\AA\ to 1500\AA, depending on the ultraviolet slope $\beta$ (F$_{\lambda} \sim \lambda^{\beta}), $ is not significant, $dm < +0.3$ if the slope is shallower than $\beta = -3$.} Uncertainties affect this conversion, such as the unknown underlying 
ultraviolet slope correction between 1500\AA\ and 2000\AA, the assumed IMF, metallicity, magnification, etc. The stellar mass limit associated to \lap\ (the entire arclet) can be relaxed to $10^4$~\msun, a value  similar to what is inferred by \citet{vanz_popiii} using the HFF photometric upper limit (a sub-component of \lap\ is discussed in Sect.~\ref{popiii_complex}).
As discussed in Sect.~\ref{sec:imaging}, there is no clear detection of \lap\ also in the \JWST\ bands redder than F200W, either individually or in their stacked version. In particular, it is worth noting that the measured line fluxes from \hg, \hb\ and \oiiidoublam\ $-$ which fall in the F356W band $-$ correspond to a magnitude in the same band of $\simeq 30.4$, a value which is out of reach at the present depth, even assuming a point source morphology.

The undetected stellar counterparts of \lap\ also make the equivalent width estimates of the lines unconstrained and only lower limits can be derived. In general, possible different sizes of the stellar and nebular components might be subject to different amplification values, making the equivalent widths lens-model dependent \citep[see discussion in][]{vanz_popiii}. However, if we assume that the stellar and nebular emissions are amplified by the same factor (both originate from the same physical region),  
the equivalent widths do not depend on the lens model and correspond to $3\sigma$ lower limits of 370(2020)\AA\ rest-frame for \lya(\ha) (see Table~\ref{tab:spec}). These 
large values imply high ionizing photon production efficiency ($\xi_{ion}$),  metal poor conditions (see Sect.~\ref{very_low_Z}) and ages of a few Myr at most \citep[e.g.,][]{withers2023_canucs, maseda20, Raiter2010, Inoue2011, schaerer2002, schaerer2003}.
From the \ha\ and the (upper-limit) ultraviolet luminosities, we infer a  $2\sigma$ lower limit estimate of the ionizing photon production efficiency of log($\xi_{ion}$[erg~Hz$^{-1}$])~$>$~26 (neglecting dust attenuation and assuming no escaping ionizing photons). Such a high value resembles the one reported by \citet{maseda20} at $z\sim 4-5$ for faint high-equivalent-width \lya\ emitters with M$_{\rm UV} \simeq -16$. It is worth stressing that these limits remain quite uncertain, being the stellar counterpart still undetected.

\subsection{\lap, a hyper metal-poor system}
\label{very_low_Z}

A sample of extremely metal-poor galaxies in the local Universe (EMPG, e.g., \citealt{annibali_tosi2022NatAs}) and at moderate redshift ($z \sim 3-4$) has recently been collected by \citet[and references therein]{nishigaki2023}, with additional  candidates selected from \JWST\ observations.  
They included EMPG with  $0.01<Z<0.1$~Z$_{\odot}$  and noted a deficit of hyper metal-poor cases ($Z<0.01$~Z$_{\odot}$),  suggesting that a metallicity floor may be present at $Z \sim 0.01$~Z$_{\odot}$, particularly in the local Universe.
The search for hyper metal-poor conditions is often carried out in the high redshift Universe, where much lower mean metallicity of the intergalactic medium is expected, $Z<0.001$~Z$_{\odot}$ \citep[e.g.][]{madau_and_dickinson2014}.
Recently, \citet{curti2023} has found a shallow slope at the low-mass-end of the mass-metallicity relation at $z=3-10$, with  $\simeq 10^{7}$~\msun\ galaxies showing  12+log(O/H)~$\simeq 7.2$ (see also \citealt{yang2023_metal}), where the solar value is 12+log(O/H)$_{\odot} = 8.69\pm 0.05$ \citep[][]{asplund2009}. At fainter luminosities,  \citet{maseda2023} identified low-metallicity galaxies in ultra-deep MUSE observations at $z=3-6.7$ with $Z=(2-30)$\% Z$_{\odot}$, preferentially showing strong \lya\ emission (exceeding 120\AA\ rest-frame equivalent width). 

We employed indirect metallicity indexes, based on the strong-line method used in literature, for the estimation of gas metallicity
\citep[e.g.,][]{pagel_R23_1979, maiolino_mannucci_2019, nakajima2023, katz2022, curti2023, maseda2023, maiolino2008}, in particular
the R23~=~(\oiiidoublam\ + \oiidoublam) / \hb\ and  R3 = (\oiiiv / \hb), 
along with the O32 index as a tracer of the ionization parameter \citep[][]{kewley2002},  O32~=~(\oiiidoublam / \oiidoublam). Since the calibration of such indexes typically spans the range 12+log(O/H)~$>7$ \citep[e.g.,][]{sanders2023}, in the case of \lap\ such calibrations need to be extended to
slightly lower values.

As discussed by \citet{izotov2021_lowZ}, a common problem of the strong-line method is the dependency on the 
ionization parameter which increases the scatter of the conversion factor at low metallicity.
Figure~\ref{fig:R23} shows the photoionization model predictions of the R23 and R3 indexes down to 12+log(O/H)~$=5.5$ as a function of the ionization parameter, as calculated by \citet{nakajima2022}, along with the location of \lap\ described in this work. The inferred R23~$=0.74_{-0.19}^{+0.21}$ and R3~$=0.55_{-0.13}^{+0.14}$  implies $6<$~12+log(O/H)~$<7$, with the exact value depending on the ionization parameter.  
\citet{izotov2021_lowZ} introduced the O32 index in the oxygen abundance estimator, with the aim of taking into account the ionization parameter and reducing the scatter in the conversion at low metallicity, 12 + log(O/H)~$< 7.0$.\footnote{Eq.~5 of Izotov et al.: 12 + log10(O/H) = $0.958 \times$log10(R23 - (0.080 - 0.00078$\times$O32)$\times$O32) + 6.805;} 

\lap\ shows faint \oiiiv\ and an the absence of \oiilam, corresponding to a lower limit of O32~$ >3(6)$ at $2(1)\sigma$. Adopting the above R23 value and O32 $1\sigma$ limit for \lap, Izotov et al. Eq.~5 gives 12+log(O/H)~$< 6.3$ (or Z~$<0.004~Z_{\odot}$). This value is also consistent with those extrapolated from the photoionization models of \citet{nakajima2022} when a high-ionization parameter is considered (high \hb\ equivalent widths), as shown in Figure~\ref{fig:R23}.
Figure~\ref{fig:MZ} shows the location of  \lap\ in 
the mass-metallicity plane compared to a collection of measurements performed in the local and high redshift Universe. While the bulk of current estimates span the range of stellar masses higher than $10^{5}$ \msun\ and $Z > 0.01 Z_{\odot}$, with galaxies at the lower 
mass tail approaching the extremely metal-poor domain, \lap\ lies in the region of the hyper metal-poor systems with  $Z < 0.01 Z_{\odot}$, and among the lowest stellar mass star-forming regions probed in the first billion years of cosmic history,
breaking the low metallicity floor currently observed \citep[][]{nishigaki2023}.


\subsection{Sub-components of \lap, a Pure Line Emitter approaching the pristine stars?}
\label{popiii_complex}

So far, we have discussed the properties of the \lap\ region as it appears on the image plane.  
The strong lensing forward modeling suggests that the sub-components A and B are separated by $\sim 150$ parsec on the source plane with estimated sizes of $< 5$ and a few tens of parsecs, respectively. The comparison between the spectra extracted from components A, B and the arclet \lap\ is reported in Appendix~\ref{LINES} (Figure~\ref{3compo}). There are two facts emerging from the NIRSpec data: (1) at positions A1,2 all the lines are detected (including weak \oiiiv), while (2) only the Balmer lines are currently measured at the locations B1,2. The faintness of \oiiiv\ at B1,2 would imply an even more severe deficit of oxygen in B. In this case, the conversion to metallicity used in the previous section based on O32 cannot be used as the O32 index is not defined. Moreover, when extracting individual spectra, the overall SNR decreases, especially for the \hb\ line. We therefore rely on the \ha\ emission from B1+B2 and adopt the ratio \ha/\hb~$\simeq 2.8$ (case B) to infer the metallicity. The 
inferred R3~$< 0.2$ and R23~$<0.4$ (at $1\sigma$) would correspond to 12+log(O/H)~$< 6$ (or $Z < 0.002~Z_{\odot}$) for component B, assuming the upper evelope of the ionization parameter grid (see red arrows in Figure~\ref{fig:R23}).
The \ha\ emissions at B1,2 are slightly elongated but still nucleated, and the magnitude limits calculated in the stacked \JWST\ image (F115W, F150W and F200W) at those locations provide even fainter limits, $m_{2000}=32.4$ at $2\sigma$ within a circular aperture of $0.12''$ diameter (corresponding to $3~\times$~FWHM in F115W and  $2~\times$~FWHM in F200W of NIRCam imaging). The magnification for each of the mirrored images B1,2 is $\mu_{tot}\simeq 100$, which implies an absolute magnitude M$_{2000}$ fainter than $-9.4$ (or intrinsic $m_{2000} > 37.4$). This limit would correspond to a stellar mass of the order of (or smaller than) $10^{3}$~\msun, under the same assumptions as in the previous section. Interestingly, the above magnitude corresponds also to the expected magnitude of a single 1000~\msun\ PopIII star \citep[][]{windhorst2018,Park2023} placed at $z=7$. 

Though still speculative, component B (see the red star symbol in Figure~\ref{fig:MZ}) might be the lowest metallicity portion of \lap. 

Finally, since \ha\ traces the same gas that produces \lya, we would expect \lya\ emission also at locations B1,2, where \ha\ is detected. However, there is an apparent deficit of \lya\ emission at those positions, at least at the available depth.
On the other hand, \lya\ emission is present in the central knot, A1,2, where also \oiiiv\ and \ha\ are detected (see Figure~\ref{fig:cutouts}). The resonant nature of the \lya\ line \citep[e.g., ][]{dijkstra2014} and the  geometry of the emitting regions offer a possible explanation.
Assuming that the IGM attenuation of \lya\ is the same for A1,2 and B1,2 and that there is no dust absorption, the deficit can be ascribed to different spatial distributions of the \lya\ and \ha\ emitting regions, with \lya\  subjected to radiative transfer processes and eventually emerging from a region that is larger than the one emitting \ha\ (which is not resonant). This would imply that \lya\ is subjected to lower 
magnification than \ha, preventing us from detecting it. 

Another intriguing possibility for the lack of continuum emission is that  component B is undergoing \ha\ emission from recombination on the surface of a self-shielding system (often referred to as fluorescence)
induced by the escaping ionizing radiation from component A, in which a very metal-poor stellar complex (whose stellar component is traced by faint \oiiiv) with M$_{\rm UV} \simeq -10$ (or fainter) acts as an efficient ionizer. Also in this case the \lya\ emission might be attenuated by radiative transfer and magnification effects as described above.
This would be the first indirect probe of escaping ionizing radiation in the reionization epoch at small spatial scales (\citealt{ribas2017ApJ84119, ribas2017ApJ84611, rauch2011} and \citealt{runnholm2023} for a similar study in the local Universe). Also in this scenario, the equivalent widths of the Balmer lines at B1,2 would be formally infinite, losing their physical meaning. It is worth stressing that the still missing stellar component (from \JWST\ imaging) and the location of A1,2 on the critical line (formally with $\mu_{tot}>500$), make any further detailed analysis challenging. Such possible scenarios and an in-depth study of component A (A1,2) will require higher SNR and will be performed in a future work.

\section{Conclusions}

In this work, we present \JWST\ follow-up observations of an extremely faint, highly magnified \lya\ arclet, originally identified at $z=6.639$ with HFF and VLT/MUSE deep observations, as a possible region hosting extremely metal-poor stars \citep[][]{vanz_popiii}. \JWST\ spectroscopic data reveal new key information on the nature of this source, corroborating the evidence for the most metal poor star-forming complex currently known, observed at an epoch of 800 Myrs after the Big Bang. Our results can be summarized as follows:  

\noindent (1)
\JWST/NIRSpec IFU observations confirm the redshift of the underlying forming system, $z=6.639$, by means of hydrogen Balmer lines, \hg, \hb\ and \ha, with a remarkably faint oxygen line, \oiiidoublam, the only metal line detected, with a ratio \oiiiv/\hb~$= 0.55 \pm 0.15$.
The flux ratios of the Balmer lines are consistent (within $2\sigma$) with the case B recombination theory, suggesting negligible dust extinction. The same case B and the ratio \ha/\lya~$\simeq 5$ imply that part of the \lya\ line is attenuated by circum-galactic or intergalactic neutral gas or is escaping on larger scales.

\noindent (2)
No significant stellar counterpart is detected in the stacked \JWST/NIRCam, NIRISS, and Hubble images, down to a UV magnitude $m_{2000}\simeq 30.4$ at $2\sigma$ level, corresponding to an intrinsic magnitude $m_{2000} > 35.8$ (or fainter than M$_{2000}=-11$). Such a low luminosity implies a stellar mass $\lesssim 10^{4}$~\msun, assuming no dust extinction and an instantaneous burst scenario.
This is currently the faintest confirmed star-forming complex during the reionization era.

\noindent (3)
The deficiency of metal lines of \lap\ implies an extremely low metallicity, 12+log(O/H)~$< 6.3$ ($Z<0.004 Z_{\odot}$).
With such a low metallicity and the above upper limit on the stellar mass, \lap\ breaks the metallicity floor ($Z\gtrsim 0.01\, Z_\odot$) observed in a variety of systems in the local and distant Universe, thus entering the hyper metal-poor regime ($Z<0.01~Z_{\odot}$) and approaching the properties expected for a pristine star-forming region. A possible detection of \heii\ emission 
remains tentative and will require further exploration.

\noindent (4) Based on a high-precision strong lensing model of MACS~J0416, the highly elongated nebular-line morphology of 
\lap, straddling the critical line at $z=6.64$, can be reproduced with two components A and B spanning $\sim 300$ pc on the source plane. \lap-B is fainter than M$_{\rm UV} = -10$, with undetected metal lines, likely lying at the lowest stellar mass and metallicity of the mass-metallicity relation (Figure~\ref{fig:MZ}). \lap-A is subject to extreme magnification ($\mu\gg 100$), falling on the critical line, appears rather compact, and shows faint \oiii\ emission. 
The nature of \lap-A is still 
uncertain and will be studied in future work.
The possibility that the Balmer emission in B is induced by escaping ionizing radiation coming from component A is also an option, and, in such a case, the first indirect probe of a transverse escaping ionizing radiation during reionization.  

Overall, \lap\ (A+B) represents an intriguing remote region of forming stars, possibly approaching the long-sought pristine zero-metal conditions. More JWST observations and future facilities, such as the ELT, will be crucial to probe key spectral lines like \heii\ of Population-III regions, reaching even fainter flux limits than what \JWST\ or 8-10m class telescopes can achieve now.

Finally, it is worth emphasizing that \lap\ was a serendipitous discovery by means of blind IFU spectroscopy obtained with VLT/MUSE through the \lya\ detection, as a pure line emitter. The results presented in this work were achievable only thanks to the IFU \JWST/NIRSpec spectroscopy, which allowed us to perform a blind two-dimensional characterization of the emission lines along the arclet and the regions across the critical line. Therefore, integral filed spectroscopy played a crucial role in this kind of science.

\begin{acknowledgements}
      We thank K. Nakajima who provided us part of the information used in Figure~\ref{fig:R23} and the extended calibrations of the indexes R23 and R3 down to 12+log(O/H) = 5.5. This work is based on observations made with the NASA/ESA/CSA 
\textit{James Webb Space Telescope} (\JWST)
and \textit{Hubble Space Telescope} (\HST). 
These observations are associated with \JWST\ GO program n.1908 (PI E. Vanzella) and GTO n.1208 (CANUCS, PI C. Willot). We acknowledge financial support through grants PRIN-MIUR 2017WSCC32 and 2020SKSTHZ. MM acknowledges support from INAF Minigrant ``The Big-Data era of cluster lensing''. MC acknowledge support from INAF Mini-grant ``Reionization and Fundamental Cosmology with High-Redshift Galaxies". EV acknowledges support from the INAF GO Grant 2022 “The revolution is around the corner: JWST will probe globular cluster precursors and Population III stellar clusters at cosmic dawn”. MB and GR acknowledge support from the Slovenian national research agency ARRS through grant N1-0238. KIC acknowledges funding from the Netherlands Research School for Astronomy (NOVA) and the Dutch Research Council (NWO) through the award of the Vici Grant VI.C.212.036. MG thanks the Max Planck Society for support through the Max Planck Research Group. This research has made use of NASA’s Astrophysics Data System, QFitsView, and SAOImageDS9, developed by Smithsonian Astrophysical Observatory.
Additionally, this work made use of the following open-source packages for Python and we are thankful to the developers of these: Matplotlib \citep{matplotlib2007}, MPDAF \citep{MPDAF2019}, PyMUSE \citep{PyMUSE2020}, Numpy \citep[][]{NUMPY2011}.
\end{acknowledgements}

%
%

\bibliographystyle{aa}
\bibliography{bib}

\begin{appendix} 

\section{Post-processing and flux calibration}
\label{postprocessing}

\begin{figure*}
\center
 \includegraphics[width=\textwidth]{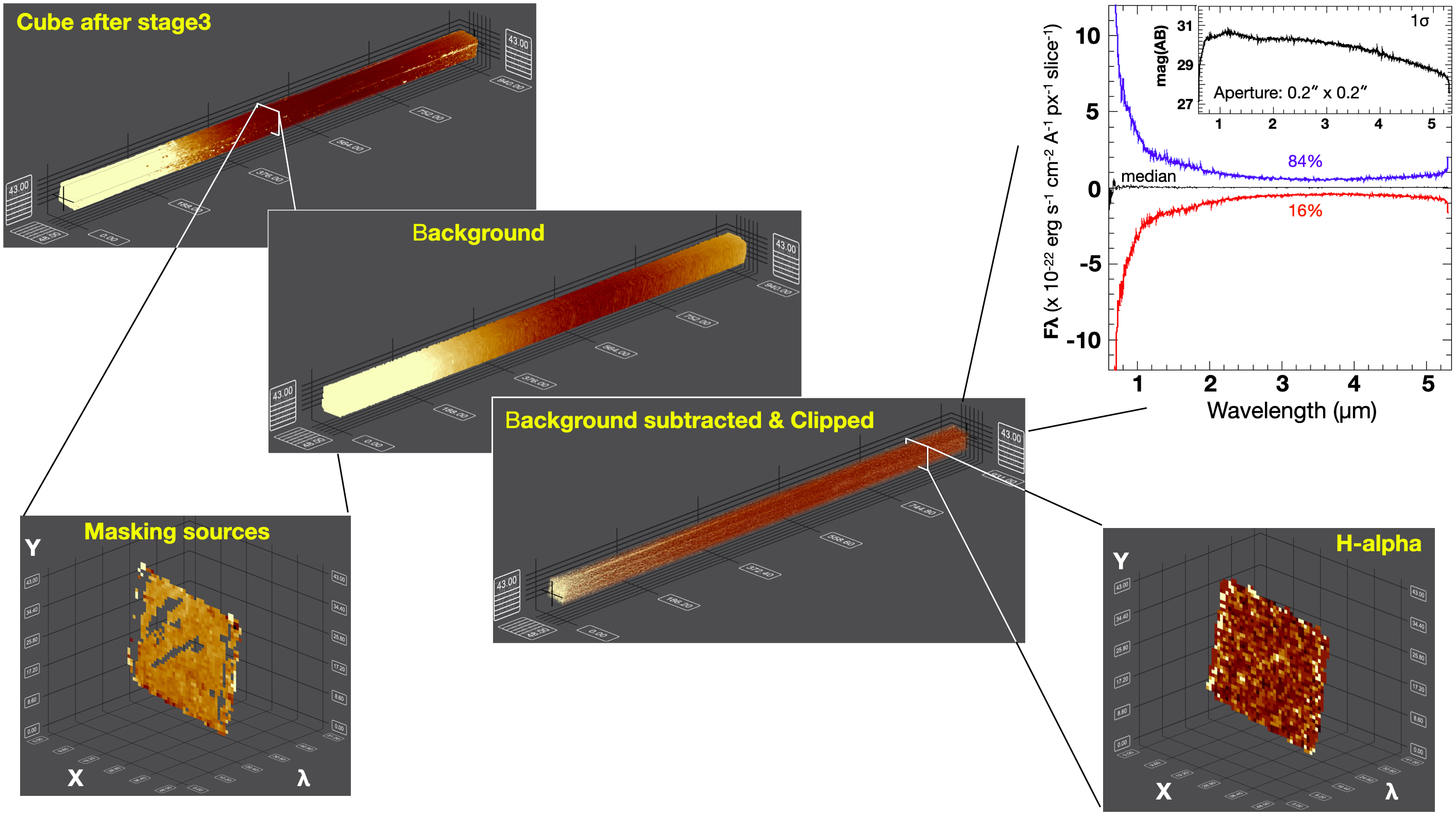}
 \caption{Scheme of the post-processing of the datacube. From top-left to lower-right the reduced cube flux calibrated in F$_\lambda$ as described in Sect.~\ref{sect:reduction}, not background subtracted. Each XY plane consists of 48 by 43 spaxels ($0.1''$/spaxel), while 940 slices span the range $0.6 - 5.3~\mu m$ along the wavelength direction, with 1 slice sampling 50\AA. In the middle panel the background is derived from the top-left with a moving median calculated on each XY plane and removing of the outliers (see text). The bottom panel shows the background-subtracted and clipped datacube. An example of masked sources and edges of the cube is shown in the bottom-left, while the bottom-right shows an extracted slice at the peak of the \ha\ line. In the top-right inset, the residual values per slice and the depth of the cube are reported (see details in the text).}
 \label{scheme}
\end{figure*}

We decided to perform the background subtraction directly from the reduced datacube rather than applying the specific procedure provided by the NIRSpec reduction  pipeline.
A limited number of sources fall in the cube field-of-view so that a large enough sky region can be used to estimate the background robustly. Specifically, the background in the datacube was derived by calculating a moving median in each XY (sky coordinates) plane, with a box of $7\times7$ 
spaxels, corresponding to $0.7'' \times 0.7''$, or $\simeq 1/20$ of the field of view ($3'' \times 3''$). The same procedure was performed by varying the window size from $3\times3$ to $11\times11$ spaxels. The few sources present in the field were masked following the NIRSpec median collapsed image and the NIRCam deep stacking, which identify the same set of sources (see Figure~\ref{fig:cutouts} and~\ref{scheme}). 
We found that median filter boxes ranging from $5\times 5$ to $9 \times 9$ spaxels provide the best results, with $7 \times 7$ providing the best compromise. 

\begin{figure*}
\center
 \includegraphics[width=\textwidth]{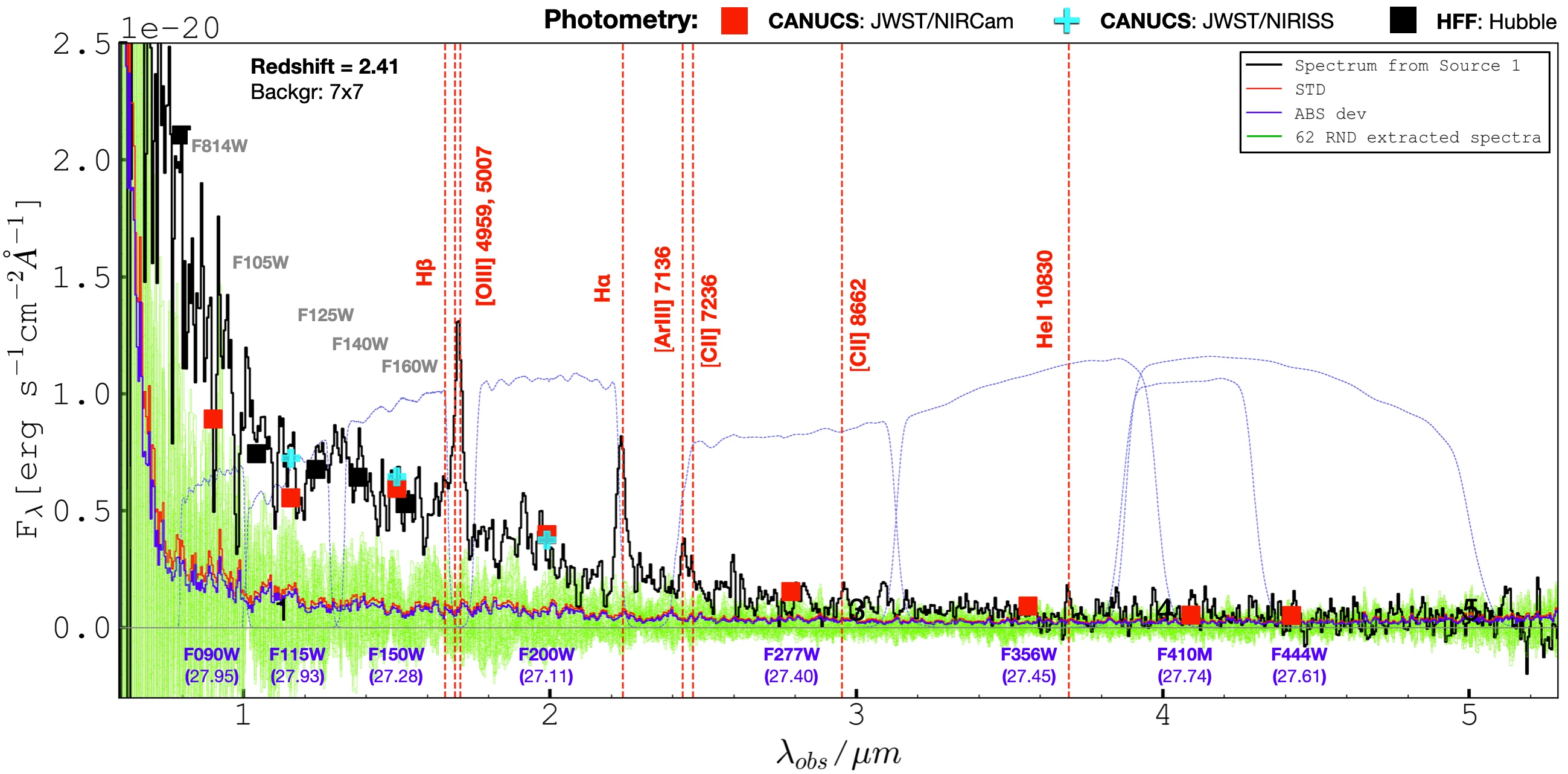}
 \caption{The one-dimensional spectrum of Source 1 at $z=2.41$ extracted from the aperture shown in Figure~\ref{fig:cutouts}. The same aperture was used to compute the \JWST/NIRcam, NIRISS, and Hubble photometry, all superimposed on this figure without applying any scaling factor. At the bottom, the filter names are indicated with the magnitudes of Source 1. The blue dashed lines show the filter throughputs. The green region shows the noise obtained from 62 spectra extracted with the same aperture size, placed randomly in the field of view. The blue and red lines show the absolute and standard deviation from the green envelope.}
 \label{Source1}
\end{figure*}

After subtracting the background cube from the original one, a cleaning procedure was applied on each XY plane by setting to zero the spaxel values deviating more than a given percentile thresholds ({\it percup} and {\it perclow}). For the arclet, we adopt percup = 98\%  and perclow 2\% after verifying that such a cut does not affect the extracted signal from \lap. 
This clipping procedure removes single spaxels and not structures. Figure~\ref{scheme} shows the median, 16\%, and 84\% percentiles calculated on each slice from 0.6 to 5.3 $\mu m$. The median is overall consistent with zero, and the lower and upper percentiles are symmetric.

Finally, the spectrum is extracted by summing up 
the corresponding spaxels values within the adopted aperture. The noise is estimated by randomly placing apertures (both of Source 1 or the arclet)  within the field of view, avoiding the targets' position.
In the case of 
the arclet, this translates into avoiding random positions along the diagonal direction of the datacube. For example, in the case of Source 1 at $z=2.41$, 62 spectra were extracted from as many random positions (shown in green in Figure~\ref{Source1}). The standard and absolute median deviations provide the 1-sigma uncertainty of the extracted spectrum. The same procedure was performed for \lap. 

The high SNR of Source 1 in the \JWST/NIRCam and NIRISS observations also provides an additional reference for flux calibration. We extracted the spectrum from the same region as the photometric aperture (see the red ellipse in Figure~\ref{fig:cutouts}). Photometry was performed with the A-PHOT tool \citep[][]{merlin19} on \JWST\ and HST data. 
Figure~\ref{Source1} shows the consistency between the photometric points and the spectrum of Source 1 without applying any correction factor. 
The flux calibration provided by the\JWST/NIRspec pipeline appears accurate at a few percent level \citep[][]{boker23}, though systematic effects may still be present on the overall normalization. We assume 
they are not larger than 20\%, as the good agreement between photometry and spectroscopy suggests. It is worth noting that this is not affecting the inferred flux ratios reported in Table~\ref{tab:spec}, especially the ratios between close (in wavelength) atomic lines.

The depth of the cleaned and flux calibrated datacube is entirely consistent with the expected one at the given integration time (Figure~\ref{scheme}, top-right), corresponding to magnitude $\simeq 30.6$ per slice at $1\sigma$ at $\lambda = 1.2~\mu m$ and within an aperture of $0.2'' \times 0.2''$ (4 spaxels).
\section{Emission lines from \lap}
\label{LINES}
The \lya, \oiiiv, and the Balmer lines, \hg, \hb\ and \ha\ are detected with SNR spanning the range $4-13$ and are shown in Figure~\ref{2dlines}, after averaging them along four slices (each slice corresponds to $\Delta\lambda=50$\AA). As expected from the lens model, two multiple images of both sides of the critical line emerge in the stacked image of Balmer lines (indicated with black circles). The \lya\  emission appears mainly located in the center, on top of the critical line, while \oiiiv\ shows a smaller extension than the Balmer lines. 

\begin{figure*}
\center
 \includegraphics[width=\textwidth]{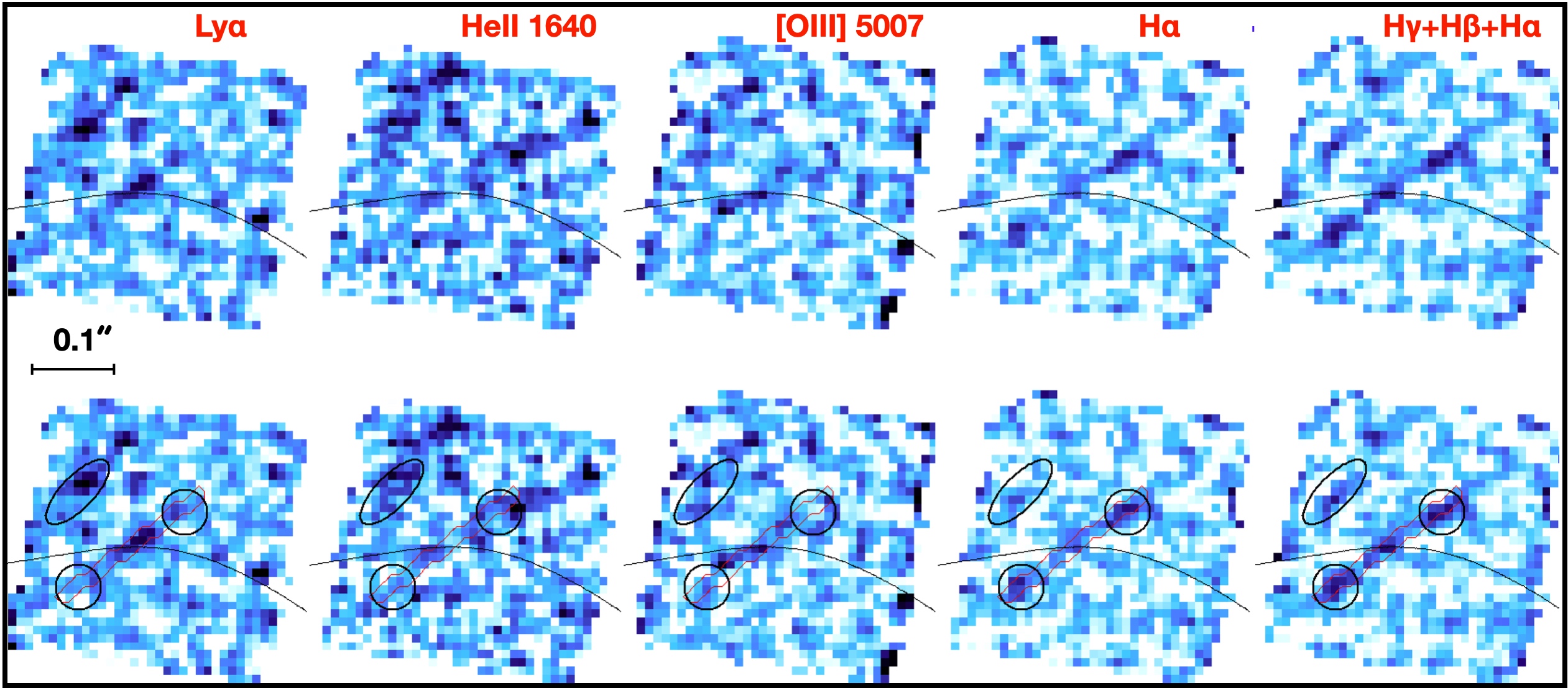}
 \caption{Two-dimensional images of the most relevant lines detected at the redshift of the arclet. From left to right, the \lya, \oiiiv, \ha\ and the sum of the Balmer lines, \hg, \hb, \ha. The first and second rows show the same images, derived after averaging three slices at the wavelength position of the lines. In the second row, the knots are labeled: 
 the black circles highlight the components B1,2 and the ellipse marks the position of source 1. The black curve outlines the critical line of the customized lens model.}
 \label{2dlines}
\end{figure*}

Figure~\ref{3compo} shows the comparison between the one-dimensional spectra extracted individually from components A1,2, B1,2 and \lap. 

\begin{figure*}
\center
 \includegraphics[width=\textwidth]{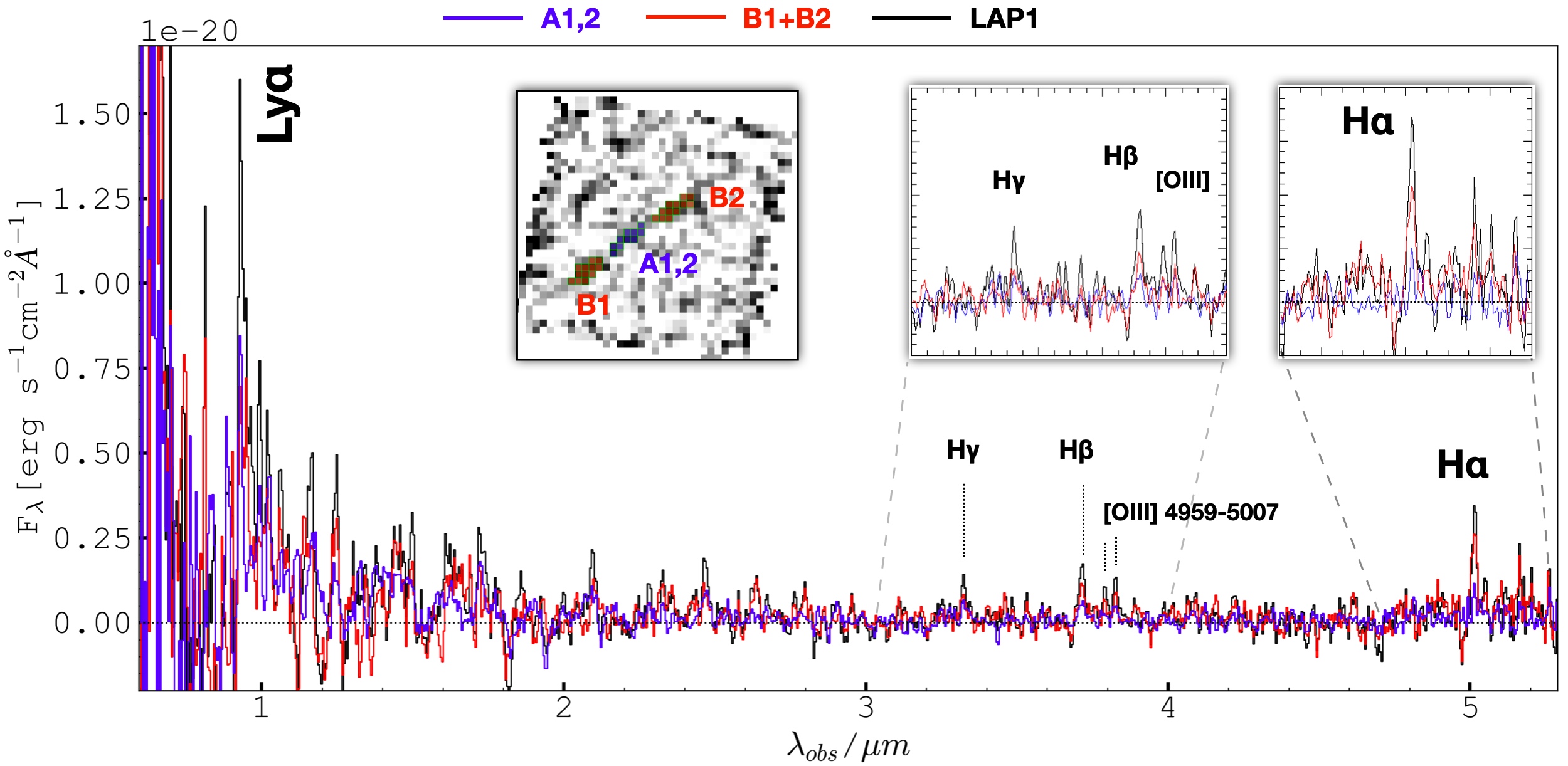}
 \caption{One-dimensional spectra extracted from individual components. The main arclet spectrum of \lap\ is shown in black, while the A1,2 and B1,2 in blue and red, respectively (as indicated in the title). The apertures from which the spectra of A1,2 and B1,2 were extracted are shown in the top-left inset in which the spaxels used are superimposed to the stacked image of Balmer lines (the same as shown in Figure~\ref{2dlines}). The top-right insets show the zoomed regions, indicated with gray-dashed lines.}
 \label{3compo}
\end{figure*}

\end{appendix}

\end{document}